\documentclass[12pt]{article}
\usepackage[T1]{fontenc}
\usepackage[utf8]{inputenc}
\usepackage{lmodern}

\usepackage{geometry}
\usepackage{setspace}
\usepackage{amsmath, amssymb, amsthm}

\usepackage{graphicx}
\usepackage[export]{adjustbox}
\usepackage{booktabs}
\usepackage{array}
\usepackage{calc}
\usepackage{longtable}
\usepackage{multirow}
\usepackage{rotating}
\usepackage{makecell}
\usepackage{threeparttable}
\usepackage{caption}
\usepackage{subcaption}
\usepackage[table]{xcolor}

\usepackage{verbatim}
\usepackage{alltt}

\usepackage[authoryear]{natbib}
\bibpunct{(}{)}{;}{a}{}{,}

\makeatletter
\@ifundefined{stlog}{%
  \newenvironment{stlog}%
    {\fontsize{10}{10}\selectfont
     \list{}{\parsep=0pt\parskip=0pt\listparindent=0pt
             \itemindent=0pt\rightmargin=0pt
             \parindent=0pt
             \topsep=8pt
             \partopsep=0pt
             \leftmargin=\leftmargini}\item\relax\verbatim}%
    {\endverbatim\endlist}%
}{}

\@ifundefined{stsyntax}{%
  \newenvironment{stsyntax}%
    {\fontsize{12}{12}\selectfont
     \list{}{\parsep=0pt\parskip=0pt\listparindent=0pt
             \itemindent=0pt\rightmargin=0pt
             \parindent=0pt\topsep=0pt
             \leftmargin=\leftmargini}\item\relax\alltt}%
    {\endalltt\endlist\vspace{3pt}}
}{}

\@ifundefined{hangpara}{%
  \def\hangpara{\hangindent=\parindent\hangafter=1\noindent}%
  \def\morehangpara{\par\hangindent=2\parindent\hangafter=1\indent}%
}{}

\@ifundefined{stresults2}{%
  \newlength{\streswidth}%
  \setlength{\streswidth}{0.5\textwidth-0.75in-\parindent}%
  \newenvironment{stresults2}%
    {\par\noindent
     \fontsize{11}{11}\selectfont
     \longtable{@{}p{1.25in}@{}p{\textwidth-1.5in-\parindent}<{\hangindent=2em\hangafter=1}@{}}}%
    {\endlongtable\par\vspace{6pt}}%
  \newcommand{\stresultsgroup}[1]{%
    \noalign{\smallskip}%
    \multicolumn{2}{@{}l@{}}{\hskip-\parindent\normalfont #1}\\[2pt]}%
}{}
\makeatother

\providecommand{\assumptionname}{Assumption}
\providecommand{\lemmaname}{Lemma}
\providecommand{\propositionname}{Proposition}
\providecommand{\remarkname}{Remark}
\providecommand{\theoremname}{Theorem}

\theoremstyle{remark}

\theoremstyle{plain}

\renewcommand{\hat}{\widehat}
\renewcommand{\tilde}{\widetilde}

\usepackage{hyperref}
\definecolor{darkblue}{rgb}{0,0,0.55}
\hypersetup{
  pdfborderstyle=,
  colorlinks=true,
  citecolor=darkblue,
  linkcolor=darkblue,
  urlcolor=darkblue
}

\usepackage{listings}
\usepackage{xcolor}
\defcitealias{mei2026nickell}{MSS26}

\title{\texttt{xtlp}: A new Stata command for panel local projection\thanks{Shu Shen, \href{mailto:shushen@link.cuhk.edu.hk}{\texttt{shushen@link.cuhk.edu.hk}}, Department of Economics, The Chinese University of Hong Kong, Hong Kong SAR, China.
Ziwei Mei, \href{mailto:ziweimei@um.edu.mo}{\texttt{ziweimei@um.edu.mo}}, Faculty of Business Administration and Asia-Pacific Academy of Economics and Management, University of Macau, Macao SAR, China.
Liugang Sheng, \href{mailto:lsheng@cuhk.edu.hk}{\texttt{lsheng@cuhk.edu.hk}}, Department of Economics, The Chinese University of Hong Kong, Hong Kong SAR, China.
Zhentao Shi, \href{mailto:zhentao.shi@cuhk.edu.hk}{\texttt{zhentao.shi@cuhk.edu.hk}}, Department of Economics, The Chinese University of Hong Kong, Hong Kong SAR, China.
Sheng acknowledges financial support from the Research Grants Council of the Hong Kong SAR (Project No.~14501821). Shi acknowledges partial financial support from the National Natural Science Foundation of China (Project No.~72425007).
}}

\author{Shu Shen \and \hspace{0.2em} Ziwei Mei \and \hspace{0.2em} Liugang Sheng \and \hspace{0.2em} Zhentao Shi}

\date{\today}

\begin{document}

\maketitle

\begin{abstract}
Panel local projection (LP) estimated with fixed effects (FEs) is widely used to evaluate the dynamic effects of economic shocks across countries or regions. However, FE estimation of panel LP models is subject to Nickell bias arising from the predictive structure of LP specifications, even when lagged dependent variables are not included in the regressors. This bias invalidates standard inference based on conventional \(t\)-statistics. We introduce \texttt{xtlp}, a new Stata command that implements panel LP with split-panel jackknife (SPJ) bias correction. It removes the Nickell bias and restores the validity of standard statistical inference. The command accommodates a broad range of specifications, including different transformations of the dependent variable, additional control variables, and flexible variance estimators. We illustrate \texttt{xtlp} through empirical applications that examine the effects of financial crises on output.

\vspace{0.5em}

\noindent \textbf{Keywords:} \texttt{xtlp}, local projections, panel data, Nickell bias, split-panel jackknife, impulse response function
\end{abstract}

\newpage
\section{Introduction}

Impulse response functions (IRFs) are essential in characterizing dynamic patterns in macroeconomic and financial systems. Since the seminal work by \citet{jorda2005estimation}, local projection (LP) has been widely used to estimate IRFs thanks to its robustness to model misspecification and easy implementation. With the growing availability of macroeconomic and financial panel data, LP is extended from time series to panel data with fixed effects (FE), capturing cross-sectional heterogeneity. The literature applies panel LP to study the economic and financial impacts of events such as financial crises, monetary and fiscal policy interventions, political shocks, and pandemics \citep{cerra2008growth,romer2017new,acemoglu2019democracy,jorda2022longer,gilchrist2022sovereign}. A comprehensive review of the LP method and its applications is available in \citet{jorda2025local}. 

Most panel LP applications use the familiar FE estimator, where the within-group transformation eliminates unobserved heterogeneity. However, statistical inference based on the FE estimator suffers from size distortion caused by the Nickell bias \citep{nickell1981biases}. The Nickell bias arises because the within transformation induces correlation between the transformed lagged dependent variable and the transformed error term, generating a finite-\(T\) bias that can be substantial under persistent dynamics. This issue is well known in dynamic panel vector autoregressions (VARs), whose underlying systems explicitly contain lagged dependent variables \citep{holtz1988estimating,hahn2002asymptotically,Arellano2003}.

Beyond panel AR models, the Nickell bias can arise in panel LP even without lagged dependent variables, which is far less recognized. \citet[MSS26 hereafter]{mei2026nickell} are the first to point out the omnipresence of the Nickell bias, cautioning that the use of the FE estimator for panel LP can cause misleading results. Valid inference for the IRFs $(\beta^{(h)})_{h=0}^H$ measuring the dynamic impact of a shock $s_{i,t}$ (e.g.~a financial crisis) on the $h$-period ahead outcome variable $y_{i,t+h}$ (e.g.~logarithmic real GDP) calls for an effective procedure to correct the Nickell bias.

The literature commonly addresses the Nickell bias using instrumental-variable (IV) estimators 
or analytical bias corrections
. However, the IV methods are prone to the weak-IV and many-IV problems, while the analytical methods require the availability of closed-form formulas. These two categories of methods therefore require case-by-case adjustment, undermining the flexibility and robustness of LP. An alternative is the split-panel jackknife (SPJ) \citep{Dhaene2015, chudik2018half}, which effectively removes the Nickell bias when the number of cross-sections and the time dimension share the same order. SPJ is easily implemented and flexible for model specifications, as it is a linear combination of three FE estimators. 

With these considerations, \citetalias{mei2026nickell} apply SPJ to remove the leading Nickell bias in panel LP regressions. In addition to the baseline settings with i.i.d.~cross-sections, their theory also accommodates several practically relevant specifications, including individual and time fixed effects, different sets of control variables, and both balanced and unbalanced panels. Hence, SPJ offers a simple and theoretically grounded correction that preserves the appeal of LP: researchers can estimate the horizon-by-horizon regression of interest without specifying the full dynamic system. Details of implementation and theoretical justifications for SPJ are available in Section \ref{sec_theory}.

In this article, we introduce the new Stata command \texttt{xtlp} to implement panel LP with SPJ bias correction. The command provides a user-friendly interface for estimating IRFs from panel data and is designed to be intuitive for users familiar with standard Stata panel-data commands such as \texttt{xtreg}. The \texttt{xtlp} command offers two main functionalities. The first is single-horizon estimation, in which \texttt{xtlp} estimates the impulse response at a specified horizon \(h\) using either the conventional FE estimator or the bias-corrected SPJ estimator. This mode is useful when researchers want to examine the results at a particular horizon in detail, including the point estimate, standard error, confidence interval, and other estimation output. The second functionality is multi-horizon estimation with automatic graphing. In this mode, \texttt{xtlp} estimates IRFs across a user-specified range of horizons and produces a plot of the IRF together with confidence intervals, providing a clear visualization of the dynamic effects of a shock over time.

The command supports a wide range of specifications commonly used in empirical work. Users can include either individual fixed effects or both individual and time fixed effects. For inference, \texttt{xtlp} provides a flexible \texttt{vce()} option, with syntax close to that used in familiar Stata commands such as \texttt{reghdfe} and \texttt{xtscc}, supporting conventional, heteroskedasticity-robust, one-way and two-way cluster-robust, and Driscoll--Kraay standard errors. The command is especially convenient for multi-horizon estimation, as \texttt{xtlp} automatically constructs the horizon-specific dependent variables over the horizon range. It allows users to apply alternative transformations of the dependent variable; for example, depending on the empirical question, the response variable can be specified in levels, first differences, cumulative changes, or other commonly used transformations. In addition, \texttt{xtlp} can estimate and plot IRFs for either a single shock or multiple shocks. Thus, estimation, inference, comparison across specifications, and default visualization of IRFs can be carried out within a single workflow. All of these features are illustrated through examples that provide detailed explanations of the command's features and their practical use.

\texttt{xtlp} is related to the existing Stata commands \texttt{xtspj} \citep{sun2019xtspj}, \texttt{lpirf}, \texttt{ivlpirf}, and \texttt{locproj} \citep{ugarte2023locproj}.  
The \texttt{xtspj} command implements the SPJ proposed by \citet{Dhaene2015} as a general incidental-parameter-bias correction method for maximum likelihood estimators in fixed-effects panel models. Our \texttt{xtlp} implements the SPJ correction proposed by \citetalias{mei2026nickell} for the FE estimator in panel LP. Accordingly, \texttt{xtlp} is tailored to the LP setting, with support for individual and two-way fixed effects as well as variance estimators commonly used in applied panel LP work; see Section \ref{sec.se} for details. The official Stata commands \texttt{lpirf} and \texttt{ivlpirf} and the community-contributed command \texttt{locproj} provide tools for estimating LP and producing IRFs; the former are designed for time-series LP, while \texttt{locproj} also offers a general interface for panel LP. \texttt{xtlp} fills the gap by correcting the Nickell bias that arises from FE in panel LP through an SPJ correction embedded directly in the LP workflow.

This article is organized as follows. Section 2 provides the theoretical background on the Nickell bias in panel LP, the SPJ correction, and SPJ-based estimation and inference. Section 3 describes the syntax, options, and stored results of the \texttt{xtlp} command. Section 4 presents detailed examples demonstrating the use of \texttt{xtlp}. Section 5 concludes.

\section{Theory and Methods} \label{sec_theory}

The theoretical foundations of \texttt{xtlp} are developed in \citetalias{mei2026nickell}, which establishes the presence of intrinsic Nickell bias in panel LP and proposes the SPJ as a general remedy. 
This section summarizes the results that motivate the design of the command, to make this paper self-contained. 

\subsection{The LP Model and the FE estimator}

Consider a panel dataset with $i = 1, \ldots, N$ cross-sectional units observed over $t = 1, \ldots, T$ time periods. Following \citet{jorda2005estimation}, we specify the panel LP model for horizon $h$ as:
\begin{equation}
y_{i,t+h} = \mu_i^{(h)y} + \beta^{(h)} s_{i,t} + \boldsymbol{\gamma}^{(h)\prime} \mathbf{c}_{i,t} + e_{i,t+h}^{(h)},
\label{eq:lp_z}
\end{equation}
where $y_{i,t+h}$ is the outcome variable, $s_{i,t}$ is the shock variable of interest, $\mathbf{c}_{i,t}$ is a vector of controls, $\mu_i^{(h)y}$ is the individual fixed effect, and $e_{i,t+h}^{(h)}$ is the error term. In a model with $p$ lags and covariates $\mathbf{z}_{i,t}$, the control vector $\mathbf{c}_{i,t}$ includes the lagged shock variable $(s_{i,t-1},\ldots,s_{i,t-p+1})^{\prime}$, the lagged outcome $(y_{i,t},y_{i,t-1},\ldots,y_{i,t-p+1})^{\prime}$, and the lagged covariates $(\mathbf{z}_{i,t}^{\prime},\mathbf{z}_{i,t-1}^{\prime},\ldots,\mathbf{z}_{i,t-p+1}^{\prime})^{\prime}$. At horizon $h=0$, the contemporaneous outcome $y_{i,t}$
is excluded from the control vector $\mathbf{c}_{i,t}$. 
We allow for lagged dependent variables in the control vector for generality, while we highlight that the Nickell bias pops up even if lagged dependent variables are absent.

The coefficient $\beta^{(h)}$ represents the IRF at horizon $h$, measuring the effect of a one-unit shock to $s_{i,t}$ on $y_{i,t+h}$. The LP approach estimates a separate regression for each horizon $h = 0, 1, \ldots, H$, producing a sequence of coefficients $\{\beta^{(h)}\}_{h=0}^{H}$ that trace the dynamic response. Unlike the VAR that derives IRFs from an estimated dynamic system, LP directly estimates the horizon-specific response in \eqref{eq:lp_z}, making it more robust to misspecification of the underlying dynamics. 

Let $\mathbf{x}_{i,t} = (s_{i,t}, \mathbf{c}_{i,t}^\prime)^\prime$ denote the vector of all regressors. We can rewrite \eqref{eq:lp_z} in compact form as 
\begin{equation}\label{eq: lp compact}
y_{i,t+h} = \mu_i^{(h)y} + \mathbf{x}_{i,t}^\prime \boldsymbol{\theta}^{(h)} + e_{i,t+h}^{(h)},
\end{equation}
where $\boldsymbol{\theta}^{(h)} = (\beta^{(h)}, \boldsymbol{\gamma}^{(h)\prime})^\prime$ is the vector of slope coefficients. The FE estimator is obtained by applying the within-group transformation to eliminate the individual fixed effects $\mu_i^{(h)y}$. Let $\mathcal{T}^h := \{p,p+1,\ldots,T-h\}$ denote the effective time-series time indices at horizon $h$, and  $T_h= T-h$. Let $[q]=\{1,\ldots,q\}$ denote the set of positive integers up to a natural number $q$.
For $t\in \mathcal{T}^h$, the within-group demeaned dependent variable is $\widetilde{y}_{i,t+h} = y_{i,t+h} - |\mathcal{T}^h|^{-1}\sum_{s\in\mathcal{T}^h } y_{i,s+h}$, and similarly for $\tilde{\mathbf{x}}_{i,t}$. The FE estimator of $\boldsymbol{\theta}^{(h)}$ in \eqref{eq: lp compact} is given by 
\begin{equation}
\widehat{\boldsymbol{\theta}}^{(h)\mathrm{fe}}  
= \left(\sum_{i\in[N]}\sum_{t\in\mathcal{T}^h} \tilde{\mathbf{x}}_{i,t} \tilde{\mathbf{x}}_{i,t}^\prime \right)^{-1} \sum_{i\in[N]}\sum_{t\in\mathcal{T}^h} \tilde{\mathbf{x}}_{i,t} \widetilde{y}_{i,t+h}.
\label{eq:FEest}
\end{equation}
The FE estimator of the IRF $\beta^{(h)}$, denoted by $\widehat{\beta}^{(h)\mathrm{fe}}$, is the first element of $\widehat{\boldsymbol{\theta}}^{(h)\mathrm{fe}}$ corresponding to $s_{i,t}$.

Statistical inference about the true IRF $\beta^{(h)}$ proceeds by constructing the $t$-statistic from $\widehat{\beta}^{(h)\mathrm{fe}}$ and its standard error, and comparing it with critical values from the standard normal distribution. The validity of this procedure rests on the strict exogeneity assumption, which requires the error term to be mean-independent of all observations of the regressors across all time periods. As we explain in the next subsection, however, the predictive structure of panel LP causes the strict exogeneity assumption to fail, thereby invalidating the conventional $t$-statistic when $N/T \to c > 0$.

\subsection{Nickell bias in panel LP} \label{sec.bias}

Econometricians have long recognized that the Nickell bias is omnipresent given predetermined regressors in panel data models \citep{Arellano2003}. 
\citetalias{mei2026nickell} focus on panel LP, showing that the FE estimator in panel LP inherently incurs the Nickell bias, regardless of whether lagged dependent variables are included among the regressors. This finding overturns the conventional perception that the Nickell bias arises only in dynamic panel LP with lagged dependent variables. 

Specifically, the FE estimator $\hat{\boldsymbol{\theta}}^{(h){\mathrm{fe}}}$ for panel LP is biased in the sense that 
\begin{align}
    &\mathbb{E}\left[\hat{\boldsymbol{\theta}}^{(h){\mathrm{fe}}} - \boldsymbol{\theta}^{(h)}\right] = \frac{\mathbf{b}^{(h)}}{T_h} + o\left(\frac{1}{T_h}\right) \label{eq: bias fe point}\\
    &\sqrt{NT_h}\left(\hat{\boldsymbol{\theta}}^{(h){\mathrm{fe}}} - \boldsymbol{\theta}^{(h)} - \frac{\mathbf{b}^{(h)}}{T_h}\right)
    \overset{d}{\to} \mathcal{N}( \mathbf{0} , \mathbf{\Sigma}). \label{eq: bias fe normal}
\end{align}
The bias $\mathbf{b}^{(h)}/T_h$ shifts the average of the asymptotic distribution away from zero, thereby inducing invalid inference based on standard $t$-statistics. When the horizon $h$ is fixed, the Nickell bias in the point estimate is of order $1/T$, while that in the asymptotic normal distribution is $\sqrt{NT_h}\left(\frac{\mathbf{b}^{(h)}}{T_h}\right) = O(\sqrt{N/T})$.

Table \ref{tab:FE_bias} summarizes the asymptotic behavior of the FE estimator.  When $N\to\infty$ with fixed $T$, the FE estimator $\hat{\boldsymbol{\theta}}^{(h){\mathrm{fe}}}$ is inconsistent, and the bias in the asymptotic distribution diverges. When $(N,T)\to\infty$ jointly, the bias in the point estimate in (\ref{eq: bias fe point}) vanishes, and the estimator is consistent. However, consistency of the point estimator does not guarantee valid inference. Under joint asymptotics with $N/T \to c \in (0,\infty)$, the leading case in macro-finance applications, \eqref{eq: bias fe normal} suggests that the limiting distribution of \(\sqrt{NT_h}(\hat{\boldsymbol\theta}^{(h)\mathrm{fe}}-\boldsymbol\theta^{(h)})\) has a nondegenerate mean \(\sqrt c\,\mathbf b^{(h)}\) of order $O(1)$. Therefore, the usual $t$-statistic is asymptotically centered away from zero, causing two-sided tests to over-reject the null hypothesis and distorting the coverage probabilities of confidence intervals. It is worth noting that such Nickell bias often leads to underestimation of the aftermath of financial crises; see Section 2.1 of \citetalias{mei2026nickell} for theoretical justifications in a prototype model and Section 3 of the same reference for empirical illustrations.

\begin{table}[!ht]
\centering
\caption{Asymptotic behavior of the FE estimator $\hat{\boldsymbol{\theta}}^{(h){\mathrm{fe}}}$}
\label{tab:FE_bias}
\footnotesize                         
\renewcommand{\arraystretch}{1}       
\setlength{\tabcolsep}{6pt} 
\begin{tabular}{lccc}
\toprule
Asymptotic regime & Bias in point estimate & Consistency & Bias in asymptotic distribution \\
\midrule
$N\to\infty$, $T$ fixed & $O(1)$ & No & $O(\sqrt{N/T})\to\infty  \vphantom{-\beta^{(0)}\dfrac{\sigma_{u_x}^{2}}{\sigma_{x}^{2}}\sqrt{c}\,\dfrac{1-\rho^{h}}{(1-\rho)^{2}}}$ \\
$(N,T)\to\infty$, $N/T\to c\in(0,\infty)$ & $O(1/T)\to 0$ & Yes & $O(1)$ \\
$(N,T)\to\infty$, $N/T\to 0$ & $O(1/T)\to 0$ & Yes & $O(\sqrt{N/T})\to 0 \vphantom{-\beta^{(0)}\dfrac{\sigma_{u_x}^{2}}{\sigma_{x}^{2}}\sqrt{c}\,\dfrac{1-\rho^{h}}{(1-\rho)^{2}}}$ \\
\bottomrule
\end{tabular}
\end{table}

To further illustrate the Nickell bias in finite samples, we replicate the simulation exercise of \citetalias{mei2026nickell}, where the prototype panel model consists of $y_{i,t} = \mu_i^{y(0)} + \beta^{(0)} x_{i,t} + \varepsilon_{i,t}$, and $x_{i,t} = \mu_i^x + \rho x_{i,t-1} + u_{i,t}$, with $\beta^{(0)} = -0.6$, $\rho = 0.8$, and $\varepsilon_{i,t}, u_{i,t} \stackrel{\mathrm{i.i.d.}}{\sim}\mathcal{N}(0,1)$. 
For each horizon $h$, we estimate the LP model $y_{i,t+h} = \mu_i^{y(h)} + \beta^{(h)} x_{i,t}  + e^{(h)}_{i,t+h}$ with FE and SPJ estimators and conduct inference for the IRF $\beta^{(h)}$. We focus on the case $N=50$ and $T=120$.

\begin{figure}[htbp]
    \centering
    \begin{subfigure}[b]{0.48\textwidth}
        \centering
        \includegraphics[width=\textwidth]{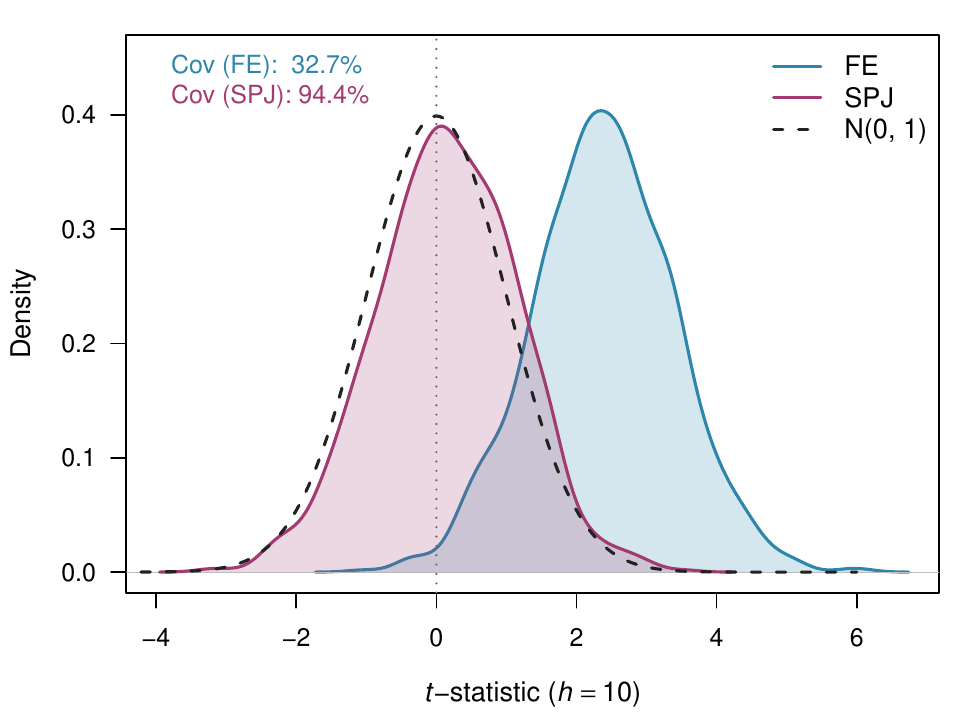}
        \caption{Density of $t$-statistics}
        \label{fig:panel_a}
    \end{subfigure}
    \hfill
    \begin{subfigure}[b]{0.48\textwidth}
        \centering
        \includegraphics[width=\textwidth]{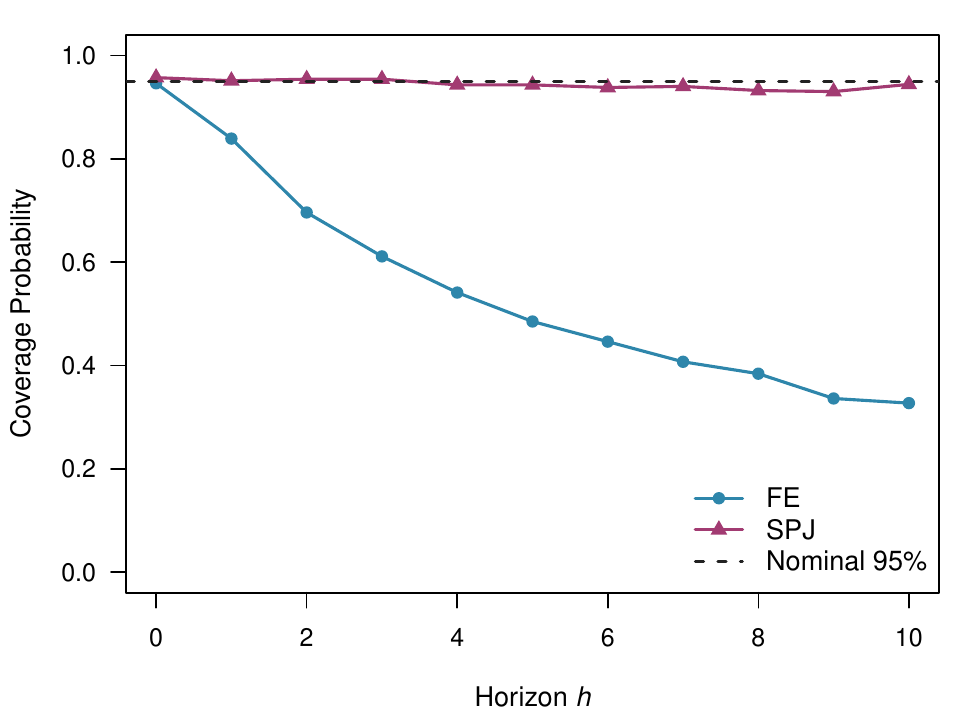}
        \caption{Coverage probabilities of confidence intervals}
        \label{fig:panel_b}
    \end{subfigure}
    \caption{Illustrative simulations for Nickell bias}
    \label{fig:illustration}
\end{figure}

Figure \ref{fig:illustration}(a) displays the density of $t$-statistics when the null hypothesis $\mathbb{H}_0:\beta^{(h)} = \beta^{(h)}_0$ is true under $h=10$, showing that the FE estimator exhibits a noticeable bias that shifts the $t$-statistic away from the standard normal distribution $\mathcal{N}(0,1)$. If a researcher ignores this bias and relies on the conventional $t$-test from the FE estimator, the type I error will be far larger than the nominal level. In contrast, the $t$-statistic from our recommended SPJ (introduced in Section \ref{subsec: SPJ}) well approximates the standard normal distribution. Figure \ref{fig:illustration}(b) exhibits the coverage probabilities of confidence intervals, highlighting that the FE estimator leads to substantial distortions in the coverage probabilities. Again, the SPJ estimator provides coverage probabilities close to the nominal level, demonstrating that SPJ effectively corrects for the Nickell bias. 

We will next formally introduce the SPJ estimator and explain how it corrects the Nickell bias for panel LP.

\subsection{Split-panel jackknife correction} \label{subsec: SPJ}

The SPJ estimator, originally developed by \citet{Dhaene2015}, is a simple and effective solution. Let $\hat{\boldsymbol{\theta}}^{(h)\text{fe}}$ be the FE estimator of \eqref{eq: lp compact}, and let $\hat{\boldsymbol{\theta}}_a^{(h)\text{fe}}$ and $\hat{\boldsymbol{\theta}}_b^{(h)\text{fe}}$ be the FE estimators from the first half $\mathcal{T}_{a}^{h}:= \{t\in\mathcal{T}^{h}: t\leq (T_{h}-p+1)/2 \}$ and second half $\mathcal{T}_{b}^{h}:=\{ t\in\mathcal{T}^{h}: t> (T_{h}-p+1)/2\}$ of the panel, respectively. The SPJ estimator is
\begin{equation}
\hat{\boldsymbol{\theta}}^{(h)\text{spj}} = 2\hat{\boldsymbol{\theta}}^{(h)\text{fe}} - \frac{1}{2}\left(\hat{\boldsymbol{\theta}}_a^{(h)\text{fe}} + \hat{\boldsymbol{\theta}}_b^{(h)\text{fe}}\right).
\label{eq:spj}
\end{equation}
The intuition behind SPJ is that the leading Nickell bias is proportional to $1/T_h$ as in \eqref{eq: bias fe point}, i.e.
\[
\mathbb{E}\!\left[\hat{\boldsymbol{\theta}}^{(h)\mathrm{fe}}-\boldsymbol{\theta}^{(h)}\right]
=
\frac{\mathbf{b}^{(h)}}{T_h}+o\!\left(\frac{1}{T_h}\right). 
\]
Furthermore, when the panel is stationary over the time dimension, each half-panel estimator has leading bias $2\mathbf{b}^{(h)}/T_h+o(1/T_h)$. Hence, the SPJ estimator satisfies
\[
\mathbb{E}\!\left[\hat{\boldsymbol{\theta}}^{(h)\mathrm{spj}}-\boldsymbol{\theta}^{(h)}\right]
= 2 \cdot \frac{\mathbf{b}^{(h)}}{T_h} - \frac{1}{2}\left(\frac{2\mathbf{b}^{(h)}}{T_h} + \frac{2\mathbf{b}^{(h)}}{T_h}\right) + o\left(\frac{1}{T_h}\right) = o\left(\frac{1}{T_h}\right),
\]
thereby removing the first-order bias. 

Define
\begin{equation}\label{eq: d star def}
\mathbf{d}_{i,t}^{*}=(\mathbf{x}_{i,t}-\bar{\mathbf{x}}_{i,b})\cdot\boldsymbol{1}\left\{ t\in\mathcal{T}_{a}^{h}\right\} +(\mathbf{x}_{i,t}-\bar{\mathbf{x}}_{i,a})\cdot\boldsymbol{1}\left\{ t\in\mathcal{T}_{b}^{h}\right\}
\end{equation} 
where $\bar{\mathbf{x}}_{i,k}=(|\mathcal{T}_{k}^{h}|)^{-1}\sum_{t\in\mathcal{T}_{k}^{h}}\mathbf{x}_{i,t}$ for $k\in\{a,b\}$, and \(\boldsymbol{1}\{\cdot\}\) denotes the indicator function. 
\citetalias{mei2026nickell} derive the asymptotic distribution of the SPJ estimator with clustering at the individual level, as $(N,T)\to\infty$, and $N/T^3\to 0$, 
\begin{equation}
\sqrt{N T_h}
\left(
\hat{\boldsymbol{\theta}}^{(h)\mathrm{spj}}
-
\boldsymbol{\theta}^{(h)}
\right)
\stackrel{d}{\to}
\mathcal{N}
\left(
\mathbf{0},
\mathbf{Q}^{-1}\mathbf{R}\mathbf{Q}^{-1}
\right), 
\label{eq:CLT_HJ}
\end{equation}
where $\mathbf{Q}$ is the Gram matrix and $\mathbf{R}$ is the asymptotic variance of $\frac{1}{\sqrt{NT_h}}\sum_{i\in[N]}\sum_{t\in\mathcal{T}^h}\mathbf{d}_{i,t}^{*}e_{i,t+h}^{(h)}$. The asymptotic distribution centered at zero allows us to apply standard inference procedures. The condition $N/T^3 \to 0$ for SPJ accommodates the leading case of $N/T\to c\in(0,\infty)$ typical in macroeconomic applications. As a practical guideline, SPJ is recommended for panels with $T\geq 30$ and $N/T\leq 10$.

\subsection{Practical Implementation of SPJ Inference}

The preceding discussion focuses on the asymptotic distribution of the SPJ estimator under individual-level clustering, thereby illustrating how the SPJ addresses the Nickell bias in LP regressions. In empirical applications, researchers may adopt alternative specifications for both the predictors and the dependence structure of the errors. These cases can be handled by suitably adapting existing econometric methods.

\subsubsection{Standard errors} \label{sec.se}

For practical implementation of inference, we need to estimate the matrices $\mathbf{Q}$ and $\mathbf{R}$ in \eqref{eq:CLT_HJ}. After we obtain consistent estimators $\hat{\mathbf{Q}}$ and $\hat{\mathbf{R}}$, the $100(1-\alpha)\%$ confidence interval for the $j$-th element in $\boldsymbol{\theta}^{(h)}$, denoted as $\theta^{(h)}_j$, can be constructed as
\[
\left[\hat{\theta}^{(h)\mathrm{spj}}_j - z_{1-\alpha/2}\sqrt{\frac{\hat s_j^2}{NT_h}},\;\hat{\theta}^{(h)\mathrm{spj}}_j + z_{1-\alpha/2}\sqrt{\frac{\hat s_j^2}{NT_h}}\right],
\]
where $\hat s_j^2$ is the $j$-th diagonal element of $\hat{\mathbf{Q}}^{-1}\hat{\mathbf{R}}\hat{\mathbf{Q}}^{-1}$, and $z_{1-\alpha/2}$ is the $(1-\alpha/2)$ quantile of the standard normal distribution, which is approximately $1.96$ when $\alpha = 0.05$. Through this approach, researchers can conduct asymptotically unbiased inference for IRFs.

The matrix $\mathbf{Q}$ is easily estimated by its sample analog
\[
\hat{\mathbf{Q}}= (NT_{h})^{-1}\sum_{i\in[N]}\sum_{t\in\mathcal{T}^h}\tilde{\mathbf{x}}_{i,t}\tilde{\mathbf{x}}_{i,t}^{\prime}.
\]
The estimator for the matrix $\mathbf{R}$ varies depending on the assumed structure of the error terms, as discussed below.

\paragraph{Homoskedastic standard errors}
Under homoskedasticity, we estimate $\mathbf{R}$ with
\[
\widehat{\mathbf{R}}_{\mathrm{homo}}
=
(NT_{h})^{-1}
\sum_{i\in[N]}\sum_{t\in\mathcal{T}^h}\mathbf{d}_{i,t}^*\mathbf{d}_{i,t}^{*\prime}
\left(\widehat{\sigma}^{\mathrm{spj}}\right)^2,
\]
where $(\widehat{\sigma}^{\mathrm{spj}})^2=(NT_{h})^{-1}\sum_{i\in[N]}\sum_{t\in\mathcal{T}^h}(\hat{e}_{i,t+h}^{(h)\mathrm{spj}})^{2}$, $\hat e_{i,t+h}^{(h)\mathrm{spj}} = \tilde y_{i,t+h} - \tilde{\mathbf{x}}_{i,t}^{\prime}\hat{\boldsymbol{\theta}}^{(h)\mathrm{spj}}$ is the SPJ residual, and $\mathbf{d}^*_{i,t}$ is defined in \eqref{eq: d star def}.

\paragraph{Heteroskedasticity-robust standard errors}
Allowing for arbitrary heteroskedasticity but no serial correlation, we estimate $\mathbf{R}$ with
\[
\widehat{\mathbf{R}}_{\mathrm{robust}}
=
(NT_{h})^{-1}\sum_{i\in[N]}\sum_{t\in\mathcal{T}^h}\mathbf{d}_{i,t}^{*}\mathbf{d}_{i,t}^{*\prime}\left(\hat{e}_{i,t+h}^{(h)\mathrm{spj}}\right)^{2}.
\]

\paragraph{One-way clustered standard errors}
Allow arbitrary within-cluster serial correlation and heteroskedasticity across clusters. For the one-way cluster case, let \(c_{i,t}^{1}\in\{1,\ldots,C^{1}\}\) denote the cluster variable for observation \((i,t)\). We then estimate $\mathbf{R}$ with
\[
\widehat{\mathbf{R}}_{\mathrm{c1}}
=
(NT_{h})^{-1}
\sum_{c^{1}\in[C^{1}]}
\left(
\sum_{(i,t):\, c_{i,t}^{1}=c^{1}}
\mathbf{d}_{i,t}^{*}
\hat{e}_{i,t+h}^{(h){\mathrm{spj}}}
\right)
\left(
\sum_{(i,t):\,c_{i,t}^{1}=c^{1}}
\mathbf{d}_{i,t}^{*}
\hat{e}_{i,t+h}^{(h){\mathrm{spj}}}
\right)^{\prime}
\]
to cluster by the variable $c_{i,t}^{1}$.

\paragraph{Two-way clustered standard errors}
For the two-way cluster case, let \(c_{i,t}^{2}\in\{1,\ldots,C^{2}\}\) denote the second cluster membership of observation \((i,t)\). We then estimate $\mathbf{R}$ with
\[
\widehat{\mathbf{R}}_{\mathrm{c2}}
=
(NT_{h})^{-1}
\sum_{c^{2}\in[C^{2}]}
\left(
\sum_{(i,t):\, c_{i,t}^{2}=c^{2}}
\mathbf{d}_{i,t}^{*}
\hat{e}_{i,t+h}^{(h){\mathrm{spj}}}
\right)
\left(
\sum_{(i,t):\,c_{i,t}^{2}=c^{2}}
\mathbf{d}_{i,t}^{*}
\hat{e}_{i,t+h}^{(h){\mathrm{spj}}}
\right)^{\prime}
\]
to cluster by the variable $c_{i,t}^{2}$. Similarly, we construct  
\[
\widehat{\mathbf{R}}_{\mathrm{c1c2}}
=
(NT_{h})^{-1}
\sum_{(c^{1},c^{2})\in[C^{1}]\times[C^{2}]}
\left(
\sum_{(i,t):\, c_{i,t}^{1}=c^{1},\, c_{i,t}^{2}=c^{2}}
\mathbf{d}_{i,t}^{*}
\hat{e}_{i,t+h}^{(h){\mathrm{spj}}}
\right)
\left(
\sum_{(i,t):\,c_{i,t}^{1}=c^{1},\, c_{i,t}^{2}=c^{2}}
\mathbf{d}_{i,t}^{*}
\hat{e}_{i,t+h}^{(h){\mathrm{spj}}}
\right)^{\prime}.
\]
Following \citet{cameron2011robust}, the two-way clustered variance estimator applies 
\[\widehat{\mathbf{R}}_{\text{TW}}=\widehat{\mathbf{R}}_{c1}+\widehat{\mathbf{R}}_{c2}-\widehat{\mathbf{R}}_{c1c2}.\]

\paragraph{Driscoll--Kraay standard errors}
Allow for both cross-sectional dependence and within-cross-sectional-unit serial correlation of unknown form. Following \citet{driscoll1998consistent}, we estimate $\mathbf{R}$ with
\[
\widehat{\mathbf{R}}_{\mathrm{DK}} := (NT_{h})^{-1}\left[\sum_{t\in\mathcal{T}^h} \mathbf{g}_{N,t} \mathbf{g}_{N,t}^\prime + \sum_{j=1}^{m(T_h)}w(j,m(T_h))\sum_{t=j+p}^{T-h}\left(\mathbf{g}_{N,t} \mathbf{g}_{N,t-j}^\prime + \mathbf{g}_{N,t-j} \mathbf{g}_{N,t}^\prime\right)\right],
\]
where $\mathbf{g}_{N,t} = \sum_{i\in[N]}\mathbf{d}_{i,t}^{*}\hat{e}_{i,t+h}^{(h){\mathrm{spj}}}$, $m(T_h)$ is the maximum lag, and $w(j,m(T_h)) = 1 - j/[m(T_h)+1]$ is the Bartlett kernel weight. Note that $\widehat{\mathbf{R}}_{\mathrm{DK}}$ consistently estimates the covariance matrix
\[
{\mathbf{R}}_{\mathrm{DK}} = (NT_h)^{-1}\sum_{i,j\in[N]}\sum_{t,s\in\mathcal{T}^h}\mathbb{E}(\mathbf{d}_{i,t}^{*}\mathbf{d}_{j,s}^{*\prime}{e}_{i,t+h}^{(h)} {e}_{j,s+h}^{(h)}).
\]
Equivalently, \(\widehat{\mathbf{R}}_{\mathrm{DK}}\) can be interpreted as the
\citet{newey1987simple} heteroskedasticity- and autocorrelation-consistent
covariance estimator applied to the time series of cross-sectional aggregate scores \(\{\mathbf{g}_{N,t}\}_{t\in\mathcal{T}^h}\).

\subsubsection{Time fixed effect}

The analysis so far has considered only cross-sectional fixed effects. In many empirical panel applications, however, researchers also include time fixed effects to control for common shocks. We therefore extend the preceding SPJ framework to a generic two-way fixed effects specification, following \citet{chudik2018half}. The extension is implemented by replacing the one-way demeaned regressor with its two-way demeaned analogue:
\[
\tilde{\mathbf{x}}_{i,t}^{\mathrm{tw}}=\mathbf{x}_{i,t}-T_{h}^{-1}\sum_{t\in\mathcal{T}^h}\mathbf{x}_{i,t}-N^{-1}\sum_{i\in[N]}\mathbf{x}_{i,t}+(NT_{h})^{-1}\sum_{i\in[N]}\sum_{t\in\mathcal{T}^h}\mathbf{x}_{i,t}
\]
with an analogous transformation for \(y_{i,t+h}\). The same SPJ formula in \eqref{eq:spj} then applies using the two-way FE estimators. Accordingly, the standard error formulas above remain valid after replacing \(\mathbf{d}_{i,t}^{*}\) with its two-way fixed effects version, defined by
\begin{align*}
\mathbf{d}_{i,t}^{* \mathrm{tw}}= & \left(\mathbf{x}_{i,t}-\bar{\mathbf{x}}_{i,b}-\frac{1}{N}\sum_{i\in[N]}(\mathbf{x}_{i,t}-\bar{\mathbf{x}}_{i,b})\right)\cdot\boldsymbol{1}\left\{ t\in\mathcal{T}_{a}^{h}\right\} \\
 & +\left(\mathbf{x}_{i,t}-\bar{\mathbf{x}}_{i,a}-\frac{1}{N}\sum_{i\in[N]}(\mathbf{x}_{i,t}-\bar{\mathbf{x}}_{i,a})\right)\cdot\boldsymbol{1}\left\{ t\in\mathcal{T}_{b}^{h}\right\}.
\end{align*}

\subsubsection{Unbalanced panel}

Although balanced panels are convenient for theoretical exposition, empirical panel data sets are often unbalanced. We therefore allow for unbalanced panels when implementing the SPJ correction, following the observation-splitting approach in Section A7 of \citet{chudik2018half}. For each cross-sectional unit \(i\), let \(T_i\) be the number of observations remaining after the relevant sample restrictions are imposed, ordered according to the time variable. The SPJ procedure divides these observations into two non-overlapping blocks along the time dimension. The cutoff is chosen as
\[
k_i=\left\lfloor \frac{T_i + 1}{2}\right\rfloor ,
\]
so that the first block contains the first \(k_i\) observations, and the second block contains the remaining \(T_i-k_i\) observations. Thus, the two blocks are of equal length when \(T_i\) is even, while the first block contains one additional observation when \(T_i\) is odd. This splitting rule ensures that each observation is used in at most one half-panel. For reliable finite-sample performance, each unit should have enough observations in both blocks.

\section{The xtlp command}

\subsection{Description}

The \texttt{xtlp} command implements panel LP with an optional SPJ bias correction, specified by the required \texttt{method()} option: \texttt{method(fe)} or \texttt{method(spj)}. As discussed in Section~\ref{sec_theory}, the FE estimator in panel LP is subject to Nickell bias, which attenuates the estimated impulse responses toward zero and invalidates standard inference. The SPJ estimator removes the first-order Nickell bias by combining FE estimates obtained from the full panel and from two time-split half-panels, while preserving the asymptotic efficiency of the FE estimator \citepalias{mei2026nickell}.

The command syntax follows standard Stata conventions and is intended to be intuitive for users familiar with panel-data commands such as \texttt{xtreg} (see [XT] \textbf{xtreg}). It supports Stata time-series operators and factor variables, allowing users to specify lags and differences with familiar notation such as \texttt{L.x} and \texttt{D.x} (see [U] \textbf{11.4.4 Time-series varlists} and [U] \textbf{11.4.3 Factor variables}). The fixed-effects structure is controlled by \texttt{fe} and \texttt{tfe}: \texttt{fe} includes individual fixed effects by default, whereas \texttt{tfe} includes both individual and time fixed effects; these options are mutually exclusive.

For inference, \texttt{xtlp} provides a flexible \texttt{vce()} option supporting five variance estimators (VCEs): conventional homoskedastic, heteroskedasticity-robust, one-way and two-way cluster-robust \citep{cameron2011robust}, and Driscoll--Kraay \citep{driscoll1998consistent}. Each estimator can be combined with the optional \texttt{ase} suboption, which suppresses the small-sample adjustment. The available VCE types are aligned with \texttt{xtreg}, \texttt{reghdfe} \citep{correia2017hdfe}, and \texttt{xtscc} \citep{hoechle2007robust} so that results are directly comparable across commands. By default, \texttt{xtlp} drops singleton groups before estimation, following \citet{correia2015singletons} and \texttt{reghdfe}, because singletons do not contribute to the within-group moment but inflate the sample size and absorbed fixed-effects degrees of freedom, thereby biasing standard errors downward. The \texttt{keepsingletons} option retains singleton groups and issues a warning.

By default, \texttt{xtlp} estimates only the contemporaneous response. Specifying multiple horizons with the \texttt{hor()} option switches the command to multi-horizon estimation, where a separate regression is estimated for each horizon and the results are stored in the IRF matrix \texttt{e(irf)}. The \texttt{ytransf()} option allows alternative transformations of the dependent variable, including levels, first differences, cumulative differences, and cumulative sums. These transformations accommodate different research questions and different interpretations of impulse responses. Multiple shock variables may be specified through the \texttt{shock()} option, allowing the dynamic effects of different shocks to be compared within the same framework. Finally, the \texttt{graph} option plots the estimated IRFs with confidence intervals, providing a compact visual summary of the dynamic effects. This allows users to inspect the overall response pattern, including its magnitude, persistence, and timing. By integrating estimation and visualization in a single workflow, \texttt{xtlp} makes it easier to interpret and present the results of dynamic panel LP.

\subsection{Syntax}

\begin{stsyntax}
xtlp \textit{depvar indepvars} [\textit{if}] [\textit{in}], \underline{m}ethod(\textit{method_name}) [fe tfe
    vce(\textit{vcetype} [, ase]) \underline{keepsin}gletons
    \underline{h}or(\textit{numlist}) \underline{ytr}ansf(\textit{transf_name}) \underline{sh}ock(\textit{integer}) \underline{g}raph]
\end{stsyntax}

\noindent \texttt{\textit{depvar}} is the dependent variable, and \texttt{\textit{indepvars}} contains the shock variable(s) and control variables. The command requires exactly one dependent variable and at least one independent variable; if more than two variables are supplied, the first is treated as \texttt{\textit{depvar}} and the remaining variables as \texttt{\textit{indepvars}}. \texttt{\textit{indepvars}} may contain factor variables (see [U] \textbf{11.4.3 Factor variables}). Both \texttt{\textit{depvar}} and \texttt{\textit{indepvars}} may contain time-series operators (see [U] \textbf{11.4.4 Time-series varlists}). A panel variable and a time variable must be set beforehand using \texttt{xtset}; see [XT] \textbf{xtset}. \texttt{method()} is required. \texttt{xtlp} drops observations with missing values in \texttt{\textit{depvar}}, \texttt{\textit{indepvars}}, the panel identifier, or the time variable. It requires Stata 14 or newer.

\subsection{Options}

\subsubsection{Estimation options}

\hangpara\texttt{method(\textit{method\_name})} specifies the estimation method: \texttt{fe} for the standard FE estimator or \texttt{spj} for the SPJ bias-corrected estimator. The SPJ estimator removes the first-order Nickell bias described in Section~\ref{subsec: SPJ}. \texttt{method()} is required.

\hangpara\texttt{fe} specifies that individual fixed effects be included in the model. This is the default when neither
\texttt{fe} nor \texttt{tfe} is specified. Individual fixed effects control for time-invariant unobserved heterogeneity across cross-sectional units. \texttt{fe} cannot be combined with \texttt{tfe}.

\hangpara\texttt{tfe} specifies that both individual and time fixed effects be included in the model. Time fixed effects control for shocks common to all cross-sectional units in a given time period. \texttt{tfe} cannot be combined with \texttt{fe}.

\hangpara\texttt{vce(\textit{vcetype} [, \texttt{ase}])} specifies the variance estimator for the coefficients. The default is \texttt{vce(unadjusted)}. \texttt{\textit{vcetype}} may be one of the following:

\morehangpara\texttt{\underline{un}adjusted} requests conventional (homoskedastic) standard errors, aligned with \texttt{xtreg, fe} and \texttt{reghdfe, vce(unadjusted)}.

\morehangpara\texttt{\underline{r}obust} requests heteroskedasticity-robust (HC1) standard errors, aligned with \texttt{reghdfe, vce(robust)}.

\morehangpara\texttt{\underline{cl}uster} \texttt{\textit{clustervar1}} [\texttt{\textit{clustervar2}}] requests one- or two-way cluster-robust standard errors.  \texttt{\textit{clustervar1}} is required, and \texttt{\textit{clustervar2}} is required for two-way clustering; at most two cluster variables are allowed. The cluster dimensions are independent of the absorbed fixed effects. One-way clustering is aligned with
\texttt{xtreg, fe vce(cluster \textit{clustvar})};
one- and two-way clustering are aligned with
\texttt{reghdfe, } \texttt{vce(cluster} \texttt{\textit{clustervars})}. Two-way clustering follows \citet{cameron2011robust}; if the estimated variance matrix is not positive semidefinite, the adjustment described therein is applied.

\morehangpara\texttt{\underline{dk}raay [lag(\#)] [, nodfadj]} requests Driscoll--Kraay standard errors \citep{driscoll1998consistent}, which are robust to cross-sectional dependence
and serial correlation and are clustered on the time variable. \texttt{lag(\#)} sets the maximum lag order; the default is $\lfloor 4(T_h/100)^{2/9}\rfloor$, following \citet{newey1994automatic}
and as in \texttt{xtscc} and \texttt{reghdfe}. If gaps are detected in the time variable, \texttt{xtlp} issues a warning but does not abort. By default, \texttt{xtlp} uses the \texttt{reghdfe} small-sample adjustment,
which accounts for degrees of freedom absorbed by the fixed effects. Specifying \texttt{nodfadj} suppresses this adjustment and yields the
\texttt{xtscc} convention. 
The suboption \texttt{nodfadj} is allowed only with \texttt{dkraay} and is mutually exclusive with \texttt{ase}.

\morehangpara The optional \texttt{ase} suboption, defined following \texttt{xtscc} but allowed with all \texttt{\textit{vcetype}}s,
returns asymptotic standard errors with no small-sample adjustment and is
mutually exclusive with \texttt{nodfadj}.
When \texttt{\textit{vcetype}} is \texttt{dkraay}, \texttt{xtlp, vce(dkraay lag(2), ase)} is aligned with
\texttt{xtscc, fe lag(2) ase}.

\hangpara \texttt{keepsingletons} requests that singleton groups not be dropped before estimation.  By default, singleton groups are dropped, following \citet{correia2015singletons} and \texttt{reghdfe}.  Retaining singleton groups may bias standard errors downward; therefore, \texttt{xtlp} issues a warning
when this option is specified.

\subsubsection{Multiple horizons options}

\hangpara\texttt{hor(\textit{numlist})} specifies the horizons at which the LP is estimated; a separate regression is estimated for each horizon $h$. This option accepts either one or two integers. If one integer $H$ is specified, such as \texttt{hor(10)}, LP is estimated for horizons 0 through $H$. If two integers $S$ and $H$ are specified, such as \texttt{hor(1 10)}, LP is estimated for horizons $S$ through $H$. The start horizon $S$ must be 0 or 1. The default is \texttt{hor(0)}, which performs a single estimation at horizon 0.

\hangpara\texttt{ytransf(\textit{transf\_name})} specifies the transformation applied to the dependent variable \texttt{\textit{depvar}} for the LP at each horizon $h$. The default is \texttt{ytransf(level)}. Let \(y_{i,t}\) denote \texttt{\textit{depvar}} below. \texttt{\textit{transf\_name}} may be one of the following:

\morehangpara\texttt{level} requests no transformation, using \(y_{i,t+h}\) as the dependent variable. This estimates the effect of a one-unit shock on the level of \(y_{i,t}\) at horizon \(h\).

\morehangpara\texttt{diff} requests the first difference, using \(y_{i,t+h} - y_{i,t+h-1}\) as the dependent variable. This estimates the effect of a one-unit shock on the change in \(y_{i,t}\) from \(t+h-1\) to \(t+h\).

\morehangpara\texttt{cmltdiff} requests the cumulative difference relative to period \(t-1\), using \(y_{i,t+h} - y_{i,t-1}\) as the dependent variable. This estimates the effect of a one-unit shock on the cumulative change in \(y_{i,t}\) from \(t-1\) to \(t+h\).

\morehangpara\texttt{cmltsum} requests the cumulative sum, using \(\sum_{k=0}^{h} y_{i,t+k}\) as the dependent variable. This transformation is useful when the dependent variable is already a first-differenced variable, such as a growth rate, so that the cumulative sum recovers the level impact over the horizon.

\hangpara\texttt{shock(\textit{integer})} specifies that the first \texttt{\textit{integer}} variables in \texttt{\textit{indepvars}} are treated as shock variables.  The remaining variables in \texttt{\textit{indepvars}} are treated as controls.  The default is \texttt{shock(1)}, so only the first variable in \texttt{\textit{indepvars}} is treated as a shock.

\hangpara\texttt{graph} requests that the estimated IRFs be graphed after estimation. This option requires multiple horizons in \texttt{hor()}.  For each shock, the graph plots point estimates and 95\% confidence intervals over the specified horizons.  If multiple shocks are specified via \texttt{shock()}, \texttt{xtlp} produces a separate graph for each shock.  The graph is generated using the Stata \texttt{twoway} command and may be customized with standard graph options.

\subsection{Stored results}

\subsubsection{Single-horizon estimation}

When \texttt{hor(0)} is specified (or implied by default), \texttt{xtlp} runs one FE or SPJ estimation and stores the following in \texttt{e()}:

\begin{stresults2}
\stresultsgroup{Scalars}
\texttt{e(N)}        & number of observations \\
\texttt{e(N\_g)}     & number of panels (individuals) \\
\texttt{e(df\_r)}    & residual degrees of freedom \\
\texttt{e(N\_clust)} & number of clusters; only with \texttt{vce(cluster ...)} \\
\texttt{e(lag)}      & Driscoll--Kraay maximum lag order; only with \texttt{vce(dkraay)} \\
\stresultsgroup{Macros}
\texttt{e(cmd)}        & \texttt{xtlp} \\
\texttt{e(cmdline)}    & command as typed \\
\texttt{e(depvar)}     & name of the dependent variable \\
\texttt{e(indepvars)}  & names of the independent variables \\
\texttt{e(vce)}        & canonical \texttt{vce()} string, e.g., \texttt{unadjusted}, \texttt{robust}, \texttt{cluster \textit{clustervar1}}, \texttt{cluster \textit{clustervar1} \textit{clustervar2}}, \texttt{dkraay lag(2)}, or \texttt{dkraay lag(2) nodfadj} \\
\texttt{e(vcetype)}    & title displayed above the standard errors: \texttt{Robust}, \texttt{Cluster}, or \texttt{Drisc/Kraay} \\
\texttt{e(properties)} & \texttt{b V} \\
\stresultsgroup{Matrices}
\texttt{e(b)} & coefficient vector \\
\texttt{e(V)} & variance--covariance matrix of the coefficients \\
\stresultsgroup{Functions}
\texttt{e(sample)} & marks the estimation sample \\
\end{stresults2}

\subsubsection{Multiple-horizon estimation}

When \texttt{hor()} specifies multiple horizons, \texttt{xtlp} estimates the model separately for each horizon $h$ in the requested range and stores one set of per-horizon results. \texttt{xtlp} stores the following:

\begin{stresults2}
\stresultsgroup{Scalars}
\texttt{e(df\_r}\textit{h}\texttt{)} & residual degrees of freedom for horizon $h$ \\
\texttt{e(lag}\textit{h}\texttt{)}    & Driscoll--Kraay maximum lag order for horizon $h$; only with \texttt{vce(dkraay)} \\
\stresultsgroup{Macros}
\texttt{e(cmd)}        & \texttt{xtlp} \\
\texttt{e(cmdline)}    & command as typed \\
\texttt{e(vce)}        & canonical \texttt{vce()} string, e.g., \texttt{unadjusted}, \texttt{robust}, \texttt{cluster \textit{clustervar1}}, \texttt{cluster \textit{clustervar1} \textit{clustervar2}}, \texttt{dkraay lag(2)}, or \texttt{dkraay lag(2) nodfadj} \\
\texttt{e(vcetype)}    & title displayed above the standard errors: \texttt{Robust}, \texttt{Cluster}, or \texttt{Drisc/Kraay} \\
\stresultsgroup{Matrices}
\texttt{e(b}\textit{h}\texttt{)} & coefficient vector for horizon $h$ \\
\texttt{e(V}\textit{h}\texttt{)} & variance--covariance matrix of the coefficients for horizon $h$ \\
\texttt{e(irf)}   & consolidated IRF results: point estimates, standard errors, and 95\% confidence interval bounds \\
\end{stresults2}

\subsection{Remark}

By default, \texttt{xtlp} applies small-sample adjustments to the reported standard errors rather than using the asymptotic variance estimators in Section~\ref{sec.se} directly. For homoskedastic and heteroskedasticity-robust standard errors, the adjustment is based on the residual degrees of freedom after accounting for the number of estimated regressors and absorbed fixed effects. For clustered and Driscoll--Kraay standard errors, \texttt{xtlp} follows the small-sample adjustment convention of \citet{correia2015singletons}: absorbed fixed effects that are nested within the clustering dimension are treated as redundant and therefore do not reduce the cluster-robust degrees of freedom. The \texttt{ase} suboption suppresses these finite-sample adjustments and reports the corresponding asymptotic standard errors.

Before estimation, \texttt{xtlp} iteratively drops singleton groups---groups with exactly one observation in the current sample after applying the within transformation. Under \texttt{tfe}, both individual and time singletons are dropped iteratively, because dropping one set can create new singletons in the other. This follows \citet{correia2015singletons} and the default behavior of \texttt{reghdfe}: singletons contribute nothing to the within-group moment but inflate the sample size and the absorbed fixed-effects degrees of freedom, which biases standard errors downward. The drop is applied separately for each horizon, because each horizon's lead defines a different estimation sample and therefore a different set of singleton groups; any horizon in which observations are dropped reports the number dropped. The \texttt{keepsingletons} option disables this behavior and issues a warning.

\section{Examples}
\label{sec:examples}

This section demonstrates the use of \texttt{xtlp} through seven examples that showcase the command's main functionalities: fixed effects, single-horizon estimation, multi-horizon estimation with graphing, transformations of the dependent variable, multiple shocks, variance estimation options, and singleton handling. The examples use data from \citet{romer2017new}, \citet{baron2021banking}, \citet{mian2017household}, and \citet{cerra2008growth}, all of which are available in the replication package of \citetalias{mei2026nickell}.

\subsection{Example 1: FE vs. SPJ with individual fixed effects}

The first example demonstrates single-horizon estimation by comparing the FE and SPJ estimators with individual fixed effects. We use data from \citet{baron2021banking} (BVX, henceforth), who study the aggregate output losses associated with banking crises. BVX construct a data set of bank equity returns for 46 advanced and emerging economies over 1870--2016. Following BVX, we use this LP specification with only individual fixed effects:
\begin{equation}
\begin{split}\Delta_{h}y_{i,t+h}=\mu_{i}^{(h)y} & +\beta^{(h)\mathrm{B}}x_{i,t}^{\mathrm{B}}+\beta^{(h)\mathrm{N}}x_{i,t}^{{\mathrm{N}}} + \boldsymbol{\gamma}^{(h)\prime} \mathbf{z}_{i,t}+e^{(h)}_{i,t+h}
\end{split}
\label{eq:bvx_t1_y}
\end{equation}
for $h=1,2,...,6$, where the dependent variable $\Delta_{h}y_{i,t+h}$ is the change in the logarithm of real GDP from year $t$ to $t+h$. The regressor $x_{i,t}^{{\mathrm{B}}}=\boldsymbol{1}\{r_{i,t}^{\mathrm{B}}\leq-30\%\}$ is a dummy variable for the bank equity crash when the bank equity index dropped more than 30\%, and similarly $x_{i,t}^{{\mathrm{N}}}=\boldsymbol{1}\{r_{i,t}^{{\mathrm{N}}}\leq-30\%\}$ is a dummy for the nonfinancial equity crash. The control variables \(\mathbf{z}_{i,t}\) include three lags of bank and nonfinancial equity crash dummies, and the contemporaneous and three lagged real GDP growth and change in credit-to-GDP. $\mu_{i}^{(h)y}$ denotes a country fixed effect at horizon \(h\).

This example estimates the effect of banking crises on cumulative output growth over a six-year horizon. In the Stata data set, the dependent variable for horizon \(h=6\), \(\Delta_{6}y_{i,t+6}\), is stored as \texttt{Fd6y}. The two shock variables \(x_{i,t}^{\mathrm{B}}\) and \(x_{i,t}^{\mathrm{N}}\) correspond to \texttt{R\_B} and \texttt{R\_N}, respectively. The remaining variables in the command are the controls collected in \(\mathbf{z}_{i,t}\). The commands below estimate Equation~\eqref{eq:bvx_t1_y} at horizon \(h=6\) twice, first using the FE estimator with \texttt{m(fe)} and then using the SPJ estimator with \texttt{m(spj)}. 

\begin{stlog}
. use bvx_t1data, clear
. keep if smp==1
. xtset ccode year

. xtlp Fd6y R_B R_N  ///
>      L1R_B L2R_B L3R_B L1R_N L2R_N L3R_N ///
>      D1y L1D1y L2D1y L3D1y D1d_y L1D1d_y L2D1d_y L3D1d_y ///
>      , m(fe)

xtlp - FE - Individual Fixed Effects

Method: Fixed effects                            Number of obs     =     2,302
Standard errors: Unadjusted
Small-sample adjustment: On

------------------------------------------------------------------------------
        Fd6y | Coefficient  Std. err.      t    P>|t|     [95% conf. interval]
-------------+----------------------------------------------------------------
         R_B |     -0.021      0.011    -1.99   0.047       -0.042      -0.000
         R_N |     -0.046      0.010    -4.53   0.000       -0.065      -0.026
       L1R_B |      0.025      0.011     2.25   0.025        0.003       0.046
       L2R_B |      0.012      0.011     1.04   0.298       -0.010       0.034
       L3R_B |      0.030      0.012     2.58   0.010        0.007       0.053
       L1R_N |     -0.024      0.010    -2.34   0.019       -0.044      -0.004
       L2R_N |     -0.030      0.010    -2.90   0.004       -0.051      -0.010
       L3R_N |     -0.020      0.011    -1.88   0.060       -0.041       0.001
         D1y |      0.266      0.071     3.74   0.000        0.127       0.405
       L1D1y |      0.195      0.073     2.68   0.007        0.052       0.338
       L2D1y |      0.276      0.071     3.88   0.000        0.136       0.416
       L3D1y |      0.405      0.068     5.97   0.000        0.272       0.538
       D1d_y |     -0.213      0.049    -4.35   0.000       -0.310      -0.117
     L1D1d_y |     -0.131      0.055    -2.37   0.018       -0.239      -0.023
     L2D1d_y |     -0.139      0.055    -2.55   0.011       -0.246      -0.032
     L3D1d_y |     -0.124      0.052    -2.39   0.017       -0.225      -0.022
------------------------------------------------------------------------------

. xtlp Fd6y R_B R_N  ///
>      L1R_B L2R_B L3R_B L1R_N L2R_N L3R_N ///
>      D1y L1D1y L2D1y L3D1y D1d_y L1D1d_y L2D1d_y L3D1d_y ///
>      , m(spj)

xtlp - SPJ - Individual Fixed Effects

Method: Split-panel jackknife                    Number of obs     =     2,302
Standard errors: Unadjusted
Small-sample adjustment: On

------------------------------------------------------------------------------
        Fd6y | Coefficient  Std. err.      t    P>|t|     [95% conf. interval]
-------------+----------------------------------------------------------------
         R_B |     -0.030      0.011    -2.65   0.008       -0.052      -0.008
         R_N |     -0.059      0.011    -5.51   0.000       -0.080      -0.038
       L1R_B |      0.030      0.012     2.59   0.010        0.007       0.052
       L2R_B |      0.011      0.012     0.92   0.355       -0.012       0.034
       L3R_B |      0.042      0.012     3.36   0.001        0.017       0.066
       L1R_N |     -0.033      0.011    -3.11   0.002       -0.054      -0.012
       L2R_N |     -0.041      0.011    -3.76   0.000       -0.063      -0.020
       L3R_N |     -0.023      0.011    -2.03   0.043       -0.045      -0.001
         D1y |      0.497      0.076     6.54   0.000        0.348       0.646
       L1D1y |      0.341      0.076     4.48   0.000        0.192       0.490
       L2D1y |      0.464      0.076     6.12   0.000        0.315       0.613
       L3D1y |      0.655      0.075     8.78   0.000        0.509       0.801
       D1d_y |     -0.117      0.064    -1.83   0.068       -0.242       0.009
     L1D1d_y |     -0.130      0.058    -2.23   0.026       -0.244      -0.016
     L2D1d_y |     -0.176      0.056    -3.13   0.002       -0.286      -0.066
     L3D1d_y |     -0.134      0.055    -2.43   0.015       -0.242      -0.026
------------------------------------------------------------------------------
\end{stlog}

This example illustrates the simplest use of \texttt{xtlp}, where the only specified option is \texttt{m()}, which selects the estimator. All other settings follow the defaults: \texttt{xtlp} includes individual fixed effects, which are country fixed effects in this application, reports conventional homoskedastic standard errors, and applies no additional adjustments. The output header summarizes the estimation setup, including the estimator, fixed-effects structure, number of observations, standard-error type, and small-sample adjustment. The coefficient table reports, for each regressor, the estimated coefficient, standard error, \(t\)-statistic, \(p\)-value, and 95\% confidence interval.

The FE and SPJ estimates differ noticeably. The coefficient on \texttt{R\_B} is \(-0.021\) under FE and \(-0.030\) under SPJ, implying that the SPJ estimate is about 43\% larger in magnitude. BVX's FE estimates show that bank equity crashes predict a sizable decline in future real GDP. The SPJ estimate points in the same direction but implies an even larger output loss. Thus, while the qualitative conclusion is unchanged, the FE estimator may understate the magnitude of the decline.

\subsection{Example 2: FE vs. SPJ with two-way fixed effects}

The second example demonstrates single-horizon estimation with two-way fixed effects, which include both country and time fixed effects. We use data from \citet{romer2017new} (RR, henceforth), who construct a narrative semiannual measure of financial distress for 24 advanced countries over 1967--2012 and examine its impact on output. We follow the LP specification used in RR's analysis:
\begin{equation}
y_{i,t+h}=\mu_{i}^{(h)y}+g_{t}^{(h)}+\beta^{(h)}x_{i,t}^{\mathrm{FD}}+\sum_{j=1}^{4}\tau^{(h)}_{j}x_{i,t-j}^{\mathrm{FD}}+\sum_{k=1}^{4}\eta^{(h)}_{k}y_{i,t-k}+e^{(h)}_{i,t+h}, \label{eq:rr}
\end{equation}
for $h=0,1,...,10$, where the dependent variable $y_{i,t+h}$ is the logarithmic real GDP, and $x_{i,t}^{{\mathrm{FD}}}$ is their new index on financial distress. $\mu_{i}^{(h)y}$ and $g_{t}^{(h)}$ indicate country and time fixed effects for each horizon $h$. Moreover, four lags of the dependent variable and the financial distress index are included as control variables.

This example estimates the effect of financial distress on real GDP at horizon \(h=10\), corresponding to five years in the semiannual data. In the Stata data set, the dependent variable for horizon \(h=10\), \(y_{i,t+10}\), is stored as \texttt{f10LNGDP}. The financial distress index \(x_{i,t}^{\mathrm{FD}}\) is stored as \texttt{CRISIS}. The variables \texttt{l1LNGDP}--\texttt{l4LNGDP} and \texttt{l1CRISIS}--\texttt{l4CRISIS} are the four lags of log real GDP and financial distress, respectively. The commands below estimate Equation~\eqref{eq:rr} at \(h=10\) using \texttt{m(fe)} and \texttt{m(spj)}, respectively.

\begin{stlog}
. use rr_f4data, clear
. xtset COUNTDUMS halfyear

. xtlp f10LNGDP CRISIS ///
>      l1LNGDP l2LNGDP l3LNGDP l4LNGDP ///
>      l1CRISIS l2CRISIS l3CRISIS l4CRISIS ///
>      , tfe m(fe)

xtlp - FE - Two-way Fixed Effects (Individual + Time)

Method: Fixed effects                            Number of obs     =     1,488
Standard errors: Unadjusted
Small-sample adjustment: On

------------------------------------------------------------------------------
    f10LNGDP | Coefficient  Std. err.      t    P>|t|     [95% conf. interval]
-------------+----------------------------------------------------------------
      CRISIS |     -0.559      0.189    -2.95   0.003       -0.931      -0.188
     l1LNGDP |      0.993      0.099    10.00   0.000        0.798       1.188
     l2LNGDP |     -0.007      0.143    -0.05   0.960       -0.287       0.273
     l3LNGDP |     -0.161      0.142    -1.13   0.258       -0.441       0.118
     l4LNGDP |     -0.009      0.101    -0.09   0.925       -0.207       0.188
    l1CRISIS |      0.248      0.253     0.98   0.327       -0.248       0.743
    l2CRISIS |      0.099      0.260     0.38   0.705       -0.411       0.608
    l3CRISIS |     -0.082      0.264    -0.31   0.758       -0.600       0.437
    l4CRISIS |     -0.176      0.216    -0.82   0.414       -0.600       0.247
------------------------------------------------------------------------------

. xtlp f10LNGDP CRISIS ///
>      l1LNGDP l2LNGDP l3LNGDP l4LNGDP ///
>      l1CRISIS l2CRISIS l3CRISIS l4CRISIS ///
>      , tfe m(spj)

xtlp - SPJ - Two-way Fixed Effects (Individual + Time)

Method: Split-panel jackknife                    Number of obs     =     1,488
Standard errors: Unadjusted
Small-sample adjustment: On

------------------------------------------------------------------------------
    f10LNGDP | Coefficient  Std. err.      t    P>|t|     [95% conf. interval]
-------------+----------------------------------------------------------------
      CRISIS |     -0.715      0.217    -3.30   0.001       -1.141      -0.290
     l1LNGDP |      1.368      0.116    11.80   0.000        1.141       1.596
     l2LNGDP |      0.044      0.160     0.27   0.785       -0.271       0.358
     l3LNGDP |     -0.224      0.161    -1.39   0.165       -0.539       0.092
     l4LNGDP |     -0.057      0.119    -0.48   0.634       -0.289       0.176
    l1CRISIS |      0.279      0.284     0.98   0.327       -0.279       0.837
    l2CRISIS |      0.108      0.292     0.37   0.711       -0.465       0.682
    l3CRISIS |     -0.064      0.297    -0.21   0.831       -0.647       0.520
    l4CRISIS |     -0.401      0.257    -1.56   0.119       -0.904       0.103
------------------------------------------------------------------------------
\end{stlog}

This example extends Example~1 to the two-way fixed-effects case by adding the option \texttt{tfe}, which includes time fixed effects in addition to the default individual fixed effects. Accordingly, the output header changes to \texttt{xtlp - FE - Two-way Fixed Effects (Individual + Time)}. For financial distress (\texttt{CRISIS}), the SPJ estimate ($-0.715$) is about 28\% larger in magnitude than the FE estimate ($-0.559$), indicating substantial Nickell bias in the FE estimator even with two-way fixed effects.

\subsection{Example 3: Estimating IRFs with multiple horizons}

The third example demonstrates multi-horizon estimation with automatic graphing. Building on Example~2, we add the horizon option \texttt{hor()} and the graph option \texttt{graph}. The command estimates the IRFs across multiple horizons and produces a plot of the IRF with confidence intervals.

\begin{stlog}
. use rr_f4data, clear
. xtset COUNTDUMS halfyear

. xtlp f0LNGDP CRISIS ///
>      l1LNGDP l2LNGDP l3LNGDP l4LNGDP ///
>      l1CRISIS l2CRISIS l3CRISIS l4CRISIS ///
>      , tfe m(fe) h(0 10) g

xtlp - FE - Two-way Fixed Effects (Individual + Time)

Method: Fixed effects
Standard errors: Unadjusted
Small-sample adjustment: On

Impulse Response Function

     Horizon |          IRF     Std. err.  95% CI Lower  95% CI Upper 
-------------+-------------------------------------------------------
           0 |     -0.29310       0.04659      -0.38441      -0.20180 
           1 |     -0.39170       0.06771      -0.52441      -0.25900 
           2 |     -0.43888       0.08541      -0.60628      -0.27148 
           3 |     -0.56780       0.09886      -0.76157      -0.37403 
           4 |     -0.62023       0.11278      -0.84128      -0.39919 
           5 |     -0.62498       0.12510      -0.87016      -0.37979 
           6 |     -0.67771       0.13796      -0.94812      -0.40731 
           7 |     -0.77602       0.15106      -1.07211      -0.47994 
           8 |     -0.70830       0.16506      -1.03181      -0.38479 
           9 |     -0.60106       0.17923      -0.95236      -0.24976 
          10 |     -0.55916       0.18941      -0.93039      -0.18792 

. xtlp f0LNGDP CRISIS ///
>      l1LNGDP l2LNGDP l3LNGDP l4LNGDP ///
>      l1CRISIS l2CRISIS l3CRISIS l4CRISIS ///
>      , tfe m(spj) h(0 10) g

xtlp - SPJ - Two-way Fixed Effects (Individual + Time)

Method: Split-panel jackknife
Standard errors: Unadjusted
Small-sample adjustment: On

Impulse Response Function

     Horizon |          IRF     Std. err.  95% CI Lower  95% CI Upper 
-------------+-------------------------------------------------------
           0 |     -0.30975       0.04872      -0.40524      -0.21426 
           1 |     -0.42009       0.07238      -0.56195      -0.27823 
           2 |     -0.46984       0.09208      -0.65033      -0.28936 
           3 |     -0.64733       0.10821      -0.85943      -0.43523 
           4 |     -0.73796       0.12421      -0.98141      -0.49450 
           5 |     -0.74347       0.13900      -1.01590      -0.47104 
           6 |     -0.79035       0.15363      -1.09147      -0.48923 
           7 |     -0.89779       0.16920      -1.22943      -0.56615 
           8 |     -0.86369       0.18642      -1.22908      -0.49830 
           9 |     -0.72165       0.20390      -1.12129      -0.32201 
          10 |     -0.71543       0.21695      -1.14064      -0.29022 
\end{stlog}

With the horizon option \texttt{h(0 10)}, \texttt{xtlp} estimates the LP specification sequentially for horizons \(h=0,\ldots,10\). The first variable in the command, \texttt{f0LNGDP}, provides the horizon-zero dependent variable, \(y_{i,t}\). Since no dependent-variable transformation is specified through \texttt{ytr()}, \texttt{xtlp} uses the default option \texttt{ytr(level)}. Therefore, for each horizon \(h\), the command automatically constructs the dependent variable \(y_{i,t+h}\), the \(h\)-period-ahead value of \(y_{i,t}\), and then estimates the corresponding horizon-specific regression. This is equivalent to manually constructing the horizon-specific dependent variables from \texttt{f0LNGDP} and repeatedly applying the single-horizon command, as shown below:

\begin{stlog}
. xtlp f0LNGDP CRISIS l1LNGDP l2LNGDP l3LNGDP l4LNGDP ///
>     l1CRISIS l2CRISIS l3CRISIS l4CRISIS, tfe m(fe)

. forvalues h = 1/10 {
.     gen f`h'LNGDP = f`h'.f0LNGDP
.     xtlp f`h'LNGDP CRISIS l1LNGDP l2LNGDP l3LNGDP l4LNGDP ///
>         l1CRISIS l2CRISIS l3CRISIS l4CRISIS, tfe m(fe)
.     xtlp f`h'LNGDP CRISIS l1LNGDP l2LNGDP l3LNGDP l4LNGDP ///
>         l1CRISIS l2CRISIS l3CRISIS l4CRISIS, tfe m(spj)
. }
\end{stlog}

Unlike manually constructing horizon-specific dependent variables and estimating each horizon in a loop, \texttt{xtlp} conveniently presents the IRFs through tables and figures. Accordingly, the command output changes from a single-horizon coefficient table to an IRF table that displays only the coefficient of the shock variable over horizons \(h=0,\ldots,10\). By default, \texttt{xtlp} treats the first variable after the dependent variable on the command line as this shock variable; in this example, it is \texttt{CRISIS}. For each horizon, the table also reports the corresponding standard error and 95\% confidence interval. The output header continues to report the estimator, fixed-effects structure, standard-error type, and small-sample adjustment.

\begin{figure}[!ht]
    \begin{center}
    \begin{subfigure}[b]{0.49\textwidth}
        \centering
        \caption{FE}
        \includegraphics[width=\textwidth]{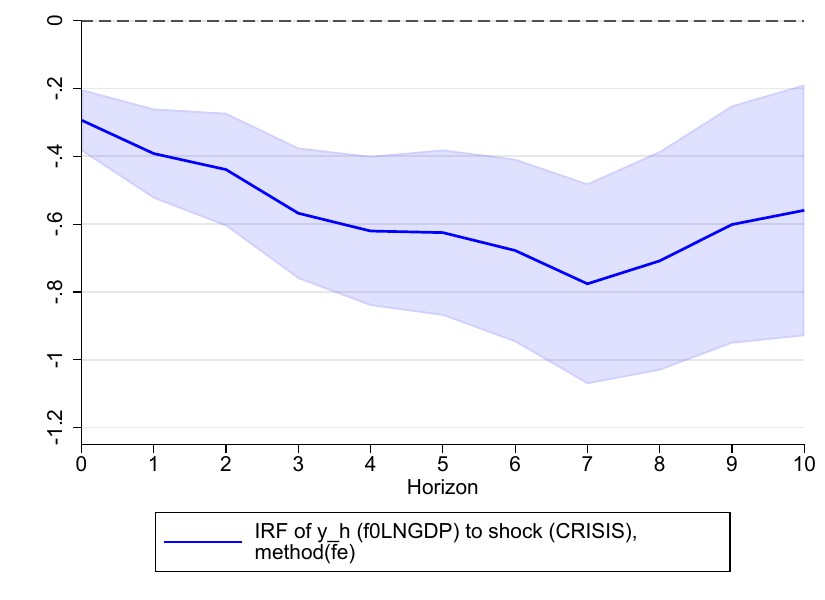} 
        \label{fig.example3a}
    \end{subfigure}
    \begin{subfigure}[b]{0.49\textwidth}
        \centering
        \caption{SPJ}
        \includegraphics[width=\textwidth]{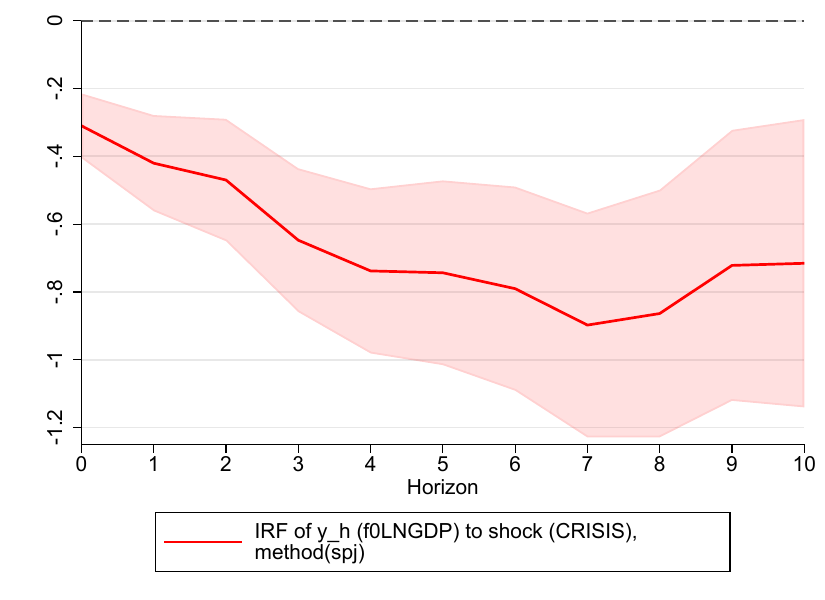} 
        \label{fig.example3b}
    \end{subfigure}
    \vspace{-0.7cm}
    \caption{Impulse responses of real GDP to financial distress}
    \label{fig.example3}
    \end{center}
    \vspace{-0.3cm}
    \footnotesize{
        \emph{Notes}: The figure shows the impulse responses of the logarithmic real GDP to financial distress, estimated using data from \citet{romer2017new}. Panel (a) shows the FE estimates, and Panel (b) shows the SPJ estimates, both with 95\% confidence intervals.
    }
\end{figure}

The graph option \texttt{g} automatically produces a plot of the IRF to the shock variable, with 95\% confidence bands, as shown in Figure~\ref{fig.example3}. The legend indicates that the blue line is the FE-based IRF of real GDP to the \texttt{CRISIS} shock, while the red line is the corresponding SPJ-based IRF. The estimated IRFs suggest a persistent negative response of output following financial crises.

Comparing the FE and SPJ estimates across horizons reveals a clear pattern: the gap between the two estimates widens at longer horizons. One year after the shock, the SPJ estimate \((-0.470)\) is about 7\% larger in magnitude than the FE estimate \((-0.439)\). Five years after the shock, the SPJ estimate \((-0.715)\) is about 28\% larger in magnitude than the FE estimate \((-0.559)\). This pattern is consistent with the theoretical analysis in Section~\ref{sec.bias} that the Nickell bias increases with the horizon. Although each estimate lies within the 95\% confidence interval of the other estimator, correcting the Nickell bias in FE remains important for valid inference. As demonstrated in Section~\ref{sec.bias}, Nickell bias can distort the centering of the FE asymptotic distribution, making the usual FE \(t\)-statistic biased and leading conventional two-sided tests of \(\mathbb{H}_0:\beta^{(h)}=0\) to over-reject and the corresponding confidence intervals to undercover.

\subsection{Example 4: Dependent variable transformation}

The fourth example demonstrates dependent-variable transformations through the \texttt{ytransf()} option. We use data from \citet{cerra2008growth}, who study the effect of currency crises on economic growth using annual panel data for 175 countries over 1965--2000. Following their analysis, we estimate using the panel LP specification:
\begin{equation}
\Delta_{h}y_{i,t+h}=\mu_{i}^{(h)y}+\beta^{(h)}x_{i,t}+\sum_{j=1}^{4}\tau_{j}^{(h)}x_{i,t-j}+\sum_{k=1}^{4}\eta_{k}^{(h)}\Delta y_{i,t-k}+e_{i,t+h}^{(h)},
\label{eq:cs}
\end{equation}
for \(h=1,2,\ldots,10\), where the dependent variable \(\Delta_{h}y_{i,t+h}\) is the change in log real GDP from year \(t-1\) to year \(t+h\), and \(x_{i,t}\) is a dummy variable indicating a currency crisis. \(\mu_i^{(h)y}\) denotes country fixed effects for each horizon \(h\). Moreover, four lags of the currency crisis dummy and four lags of real GDP growth $\Delta y_{i,t}$ are included as control variables.

In the Stata data set, real GDP growth $\Delta y_{i,t}$ is stored as \texttt{GRRT\_WB}, and the currency crisis dummy \(x_{i,t}\) is stored as \texttt{CRISIS}. The variables \texttt{l1CRISIS}--\texttt{l4CRISIS} and \texttt{l1GRRT\_WB}--\texttt{l4GRRT\_WB} are the four lags of the currency crisis dummy and real GDP growth, respectively. The option \texttt{h(1 10)} sets the horizon range.

When analyzing cumulative effects, users do not need to manually construct cumulative outcomes from \texttt{GRRT\_WB}, because the \texttt{ytr(cmltsum)} option in \texttt{xtlp} automatically constructs the cumulative sum of future real GDP growth rates. Thus, at each horizon \(h\), the transformed dependent variable corresponds to the cumulative change in log real GDP over the projection horizon, i.e., \(\Delta_{h}y_{i,t+h} = y_{i,t+h} - y_{i,t-1} = \sum_{s=0}^{h} \Delta y_{i,t+s}\). The interpretation of the IRF changes accordingly: the coefficient at horizon \(h\) represents the cumulative percentage change in GDP from period \(t\) to period \(t+h\). In the graph legend, this transformed outcome is denoted by ``cy\_h''.

\begin{stlog}
. use cs_f3data, clear
. xtset cnty obs

. xtlp GRRT_WB CRISIS ///
>      l1CRISIS l2CRISIS l3CRISIS l4CRISIS ///
>      l1GRRT_WB l2GRRT_WB l3GRRT_WB l4GRRT_WB ///
>      , fe m(fe) h(1 10) ytr(cmltsum) g

xtlp - FE - Individual Fixed Effects
(h=1: dropped 9 singleton observations)
(h=2: dropped 5 singleton observations)
(h=3: dropped 10 singleton observations)
(h=4: dropped 3 singleton observations)
(h=5: dropped 3 singleton observations)
(h=6: dropped 4 singleton observations)
(h=8: dropped 3 singleton observations)
(h=9: dropped 1 singleton observations)
(h=10: dropped 1 singleton observations)

Method: Fixed effects
Standard errors: Unadjusted
Small-sample adjustment: On

Impulse Response Function

     Horizon |          IRF     Std. err.  95% CI Lower  95% CI Upper 
-------------+-------------------------------------------------------
           1 |     -2.58895       0.31106      -3.19864      -1.97926 
           2 |     -3.38491       0.41605      -4.20038      -2.56945 
           3 |     -4.20226       0.51326      -5.20825      -3.19628 
           4 |     -4.54347       0.59566      -5.71096      -3.37597 
           5 |     -4.58089       0.65691      -5.86844      -3.29335 
           6 |     -5.13432       0.70008      -6.50647      -3.76217 
           7 |     -4.71969       0.73323      -6.15682      -3.28256 
           8 |     -4.61254       0.76871      -6.11922      -3.10586 
           9 |     -4.31107       0.80590      -5.89065      -2.73150 
          10 |     -4.50657       0.84205      -6.15698      -2.85616 

. xtlp GRRT_WB CRISIS ///
>      l1CRISIS l2CRISIS l3CRISIS l4CRISIS ///
>      l1GRRT_WB l2GRRT_WB l3GRRT_WB l4GRRT_WB ///
>      , fe m(spj) h(1 10) ytr(cmltsum) g

xtlp - SPJ - Individual Fixed Effects
(h=1: dropped 9 singleton observations)
(h=2: dropped 5 singleton observations)
(h=3: dropped 10 singleton observations)
(h=4: dropped 3 singleton observations)
(h=5: dropped 3 singleton observations)
(h=6: dropped 4 singleton observations)
(h=8: dropped 3 singleton observations)
(h=9: dropped 1 singleton observations)
(h=10: dropped 1 singleton observations)

Method: Split-panel jackknife
Standard errors: Unadjusted
Small-sample adjustment: On

Impulse Response Function

     Horizon |          IRF     Std. err.  95% CI Lower  95% CI Upper 
-------------+-------------------------------------------------------
           1 |     -2.61929       0.33682      -3.27945      -1.95913 
           2 |     -3.48836       0.45271      -4.37566      -2.60105 
           3 |     -4.78761       0.56285      -5.89079      -3.68442 
           4 |     -5.48468       0.65213      -6.76285      -4.20651 
           5 |     -6.08834       0.72733      -7.51392      -4.66276 
           6 |     -7.09202       0.78407      -8.62880      -5.55524 
           7 |     -7.39620       0.83635      -9.03544      -5.75696 
           8 |     -7.74724       0.87938      -9.47082      -6.02365 
           9 |     -7.61327       0.92486      -9.42599      -5.80055 
          10 |     -7.89262       0.96875      -9.79136      -5.99388 
\end{stlog}

Equivalently, one can manually construct the horizon-specific dependent variables from \texttt{GRRT\_WB} and repeatedly apply the single-horizon command, as illustrated below:

\begin{stlog}
. forvalues h = 1/10 {
.     gen cf`h'GRRT_WB = GRRT_WB
.     forvalues k = 1/`h' {
.         replace cf`h'GRRT_WB = cf`h'GRRT_WB + f`k'GRRT_WB
.     }
.     xtlp cf`h'GRRT_WB CRISIS l1CRISIS l2CRISIS l3CRISIS l4CRISIS ///
>          l1GRRT_WB l2GRRT_WB l3GRRT_WB l4GRRT_WB, fe m(fe)
.     xtlp cf`h'GRRT_WB CRISIS l1CRISIS l2CRISIS l3CRISIS l4CRISIS ///
>          l1GRRT_WB l2GRRT_WB l3GRRT_WB l4GRRT_WB, fe m(spj)
. }
\end{stlog}

\begin{figure}[!ht]
    \begin{center}
    \begin{subfigure}[b]{0.49\textwidth}
        \centering
        \caption{FE}
        \includegraphics[width=\textwidth]{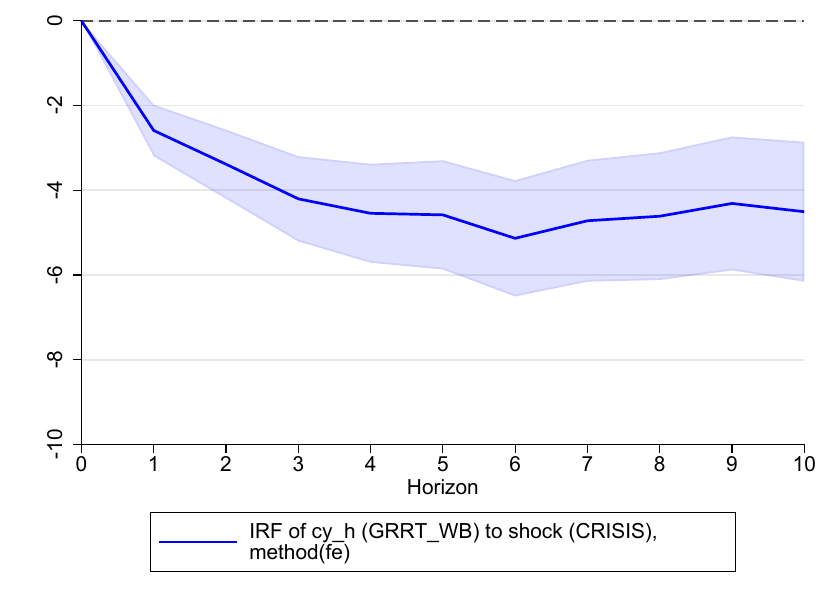} 
        \label{fig.example4a}
    \end{subfigure}
    \begin{subfigure}[b]{0.49\textwidth}
        \centering
        \caption{SPJ}
        \includegraphics[width=\textwidth]{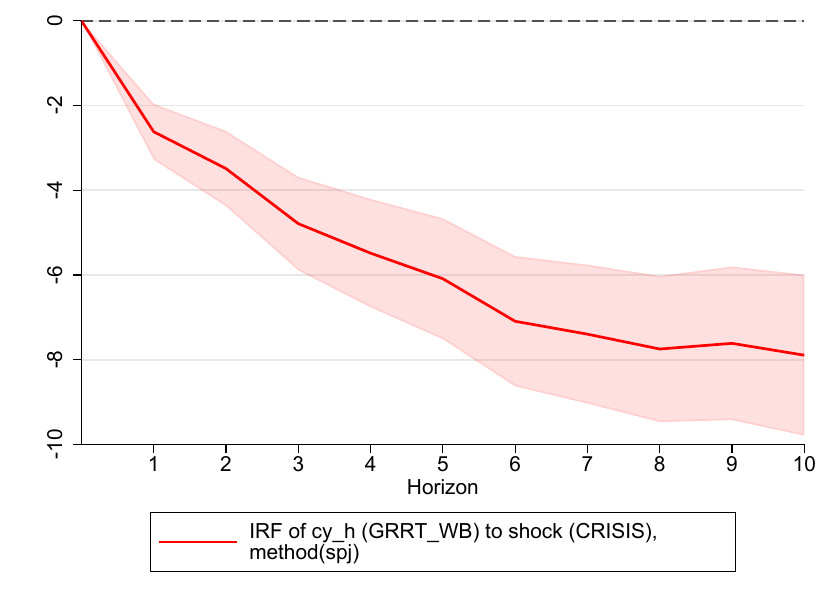} 
        \label{fig.example4b}
    \end{subfigure}
    \vspace{-0.7cm}
    \caption{Impulse responses of real GDP to currency crises}
    \label{fig.example4}
    \end{center}
    \vspace{-0.3cm}
    \footnotesize{
        \emph{Notes}: The figure shows the impulse responses of the cumulative change in log real GDP to currency crises, estimated using data from \citet{cerra2008growth}. Panel (a) shows the FE estimates, and Panel (b) shows the SPJ estimates, both with 95\% confidence intervals.
    }
\end{figure}

Figure~\ref{fig.example4} illustrates the cumulative changes in real GDP following currency crises. The most striking feature of the figure is the large discrepancy between the FE and SPJ estimates. In this example, the panel has a relatively large cross-sectional dimension \(N\) compared with the time dimension \(T\), a setting in which Nickell bias can be particularly pronounced. Consistent with this concern, the FE estimator implies a relatively moderate decline in real GDP after a currency crisis, whereas the SPJ estimator points to a much larger and more persistent cumulative output loss. Quantitatively, the FE estimates indicate an initial GDP decline of about 2.6\%, stabilizing at around 4.5\% after five years. By contrast, the SPJ estimates imply a decline exceeding 7\% after five years, suggesting that the conventional FE estimator substantially understates the economic cost of currency crises.

The divergence becomes especially pronounced at longer horizons. The FE estimate reaches a peak cumulative output loss of approximately 5.1\% around six years after the crisis, while the corresponding SPJ estimate is about 7.1\%, roughly 40\% larger. The SPJ estimates further suggest that the peak effect occurs about a decade after the crisis, with a cumulative output loss of around 7.9\%, approximately 75\% larger than the FE estimate. Moreover, for long horizons \((h>6)\), the FE estimates fall outside the 95\% confidence intervals of the SPJ estimates, indicating that the difference between the two estimators is not only economically large but also statistically meaningful. This pattern highlights the importance of bias correction in panels with large \(N\) relative to \(T\).

\subsection{Example 5: Multiple shocks}

The fifth example demonstrates the estimation of IRFs for multiple shock variables using the \texttt{shock()} option. We use data from \citet{mian2017household}, who examine the dynamics between household debt,
nonfinancial firm debt, and economic fluctuations using an unbalanced panel of 30 countries from 1960 to 2012. Following their analysis, we estimate:
\begin{equation}
y_{i,t+h}=\mu_{i}^{(h)y}+\beta^{(h)\mathrm{HH}}x_{i,t}^{\mathrm{HH}}+\beta^{(h)\mathrm{F}}x_{i,t}^{\mathrm{F}}+\sum_{j=1}^{4}\left(\tau^{(h)\mathrm{HH}}_{j}x_{i,t-j}^{\mathrm{HH}}+\tau^{(h)\mathrm{F}}_{j}x_{i,t-j}^{\mathrm{F}}\right)+\sum_{k=0}^{4}\eta^{(h)}_{k}y_{i,t-k}+e^{(h)}_{i,t+h}\label{eq:msv}
\end{equation}
for \(h=1,2,\ldots,10\), where the dependent variable \(y_{i,t+h}\) is log
real GDP at horizon \(h\). The key shock variables are the household
debt-to-GDP ratio \(x_{i,t}^{\mathrm{HH}}\) and the nonfinancial firm
debt-to-GDP ratio \(x_{i,t}^{\mathrm{F}}\). \(\mu_i^{(h)y}\) denotes country
fixed effects for each horizon \(h\). Moreover, four lags of two shock variables, together with contemporaneous and up to four-year lagged log real GDP, are included as control variables.

In the Stata data set, log real GDP \(y_{i,t}\) is stored as
\texttt{F1y}, the household debt-to-GDP ratio
\(x_{i,t}^{\mathrm{HH}}\) is stored as \texttt{L0HHD\_L1GDP},
and the nonfinancial firm debt-to-GDP ratio \(x_{i,t}^{\mathrm{F}}\) is stored
as \texttt{L0NFD\_L1GDP}. The remaining variables in the command are the control variables. The \texttt{sh(2)} option specifies that the first two variables after the dependent variable \texttt{F1y}, that is \texttt{L0HHD\_L1GDP} and \texttt{L0NFD\_L1GDP}, are treated as separate shock variables. Accordingly, the command estimates and reports separate IRFs for each shock. 

\begin{stlog}
. use msv_f2data, clear
. keep CountryCode year F1y F2y F3y F4y F5y F6y F7y F8y F9y F10y ///
>      L0HHD_L1GDP L1HHD_L1GDP L2HHD_L1GDP L3HHD_L1GDP L4HHD_L1GDP ///
>      L0NFD_L1GDP L1NFD_L1GDP L2NFD_L1GDP L3NFD_L1GDP L4NFD_L1GDP ///
>      L0y L1y L2y L3y L4y
. xtset CountryCode year

. xtlp F1y L0HHD_L1GDP L0NFD_L1GDP ///
>      L1HHD_L1GDP L2HHD_L1GDP L3HHD_L1GDP L4HHD_L1GDP ///
>      L1NFD_L1GDP L2NFD_L1GDP L3NFD_L1GDP L4NFD_L1GDP ///
>      L0y L1y L2y L3y L4y ///
>      , fe m(spj) h(0 9) sh(2) g

xtlp - SPJ - Individual Fixed Effects
(h=5: dropped 1 singleton observations)
(h=6: dropped 1 singleton observations)
(h=8: dropped 1 singleton observations)

Method: Split-panel jackknife
Standard errors: Unadjusted
Small-sample adjustment: On

Impulse Response Function for shock #1 (L0HHD_L1GDP)

     Horizon |          IRF     Std. err.  95% CI Lower  95% CI Upper 
-------------+-------------------------------------------------------
           0 |      0.20275       0.04795       0.10877       0.29672 
           1 |      0.25405       0.07832       0.10053       0.40756 
           2 |      0.20526       0.09963       0.00999       0.40053 
           3 |      0.01365       0.11906      -0.21970       0.24701 
           4 |     -0.19164       0.12901      -0.44450       0.06123 
           5 |     -0.40943       0.13629      -0.67655      -0.14232 
           6 |     -0.55770       0.14602      -0.84390      -0.27149 
           7 |     -0.49004       0.16134      -0.80626      -0.17382 
           8 |     -0.42783       0.17372      -0.76832      -0.08733 
           9 |     -0.55530       0.18987      -0.92744      -0.18316 

Impulse Response Function for shock #2 (L0NFD_L1GDP)

     Horizon |          IRF     Std. err.  95% CI Lower  95% CI Upper 
-------------+-------------------------------------------------------
           0 |     -0.07209       0.02078      -0.11281      -0.03137 
           1 |     -0.12147       0.03297      -0.18608      -0.05686 
           2 |     -0.11850       0.04268      -0.20215      -0.03484 
           3 |     -0.12916       0.05071      -0.22856      -0.02976 
           4 |     -0.13603       0.05814      -0.24998      -0.02209 
           5 |     -0.08856       0.06117      -0.20845       0.03133 
           6 |     -0.06685       0.06478      -0.19382       0.06012 
           7 |     -0.01647       0.07101      -0.15565       0.12272 
           8 |     -0.03182       0.07324      -0.17537       0.11174 
           9 |      0.02382       0.07568      -0.12452       0.17217 
\end{stlog}

\begin{figure}[!ht]
    \begin{center}
    \begin{subfigure}[b]{0.49\textwidth}
        \centering
        \caption{Household debt: SPJ}
        \includegraphics[width=\textwidth]{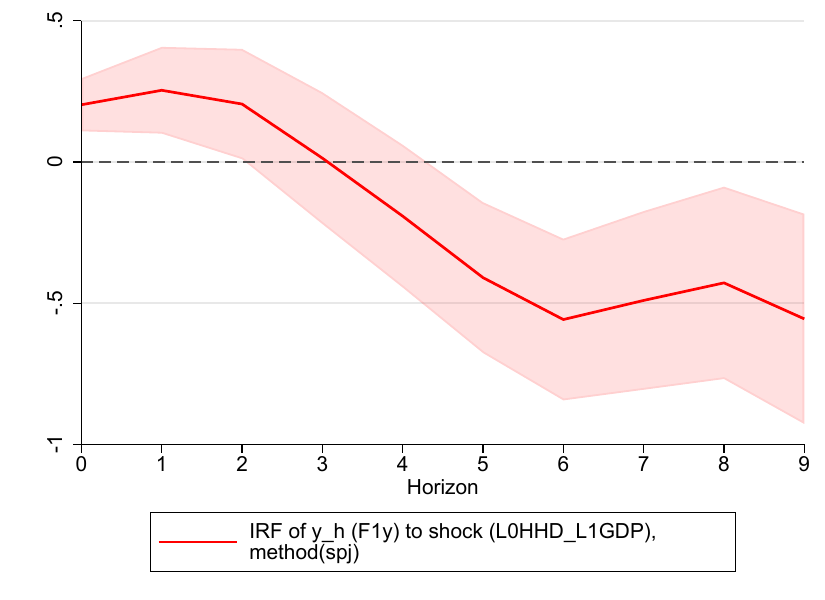} 
        \label{fig.example5_spj_1}
    \end{subfigure}
    \begin{subfigure}[b]{0.49\textwidth}
        \centering
        \caption{Nonfinancial firm debt: SPJ}
        \includegraphics[width=\textwidth]{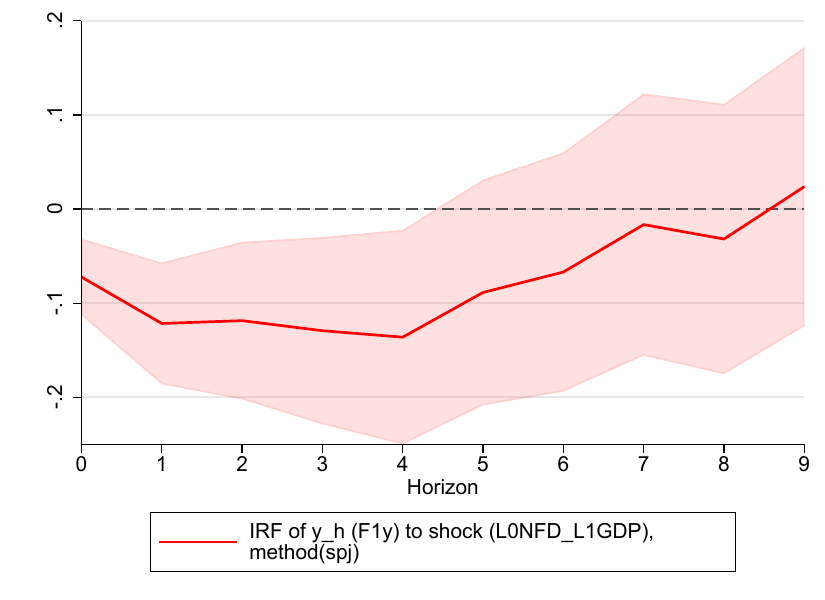} 
        \label{fig.example5_spj_2}
    \end{subfigure}
    \vspace{-0.7cm}
    \caption{Impulse responses of real GDP to household debt and nonfinancial firm debt}
    \label{fig.example5}
    \end{center}
    \vspace{-0.3cm}
    \footnotesize{
        \emph{Notes}: The figure shows the impulse responses of the logarithmic real GDP to household debt and nonfinancial firm debt, respectively, estimated using data from \citet{mian2017household}. Panels (a) and (b) show the SPJ estimates for the two shocks, both with 95\% confidence intervals.
    }
\end{figure}

The \texttt{graph} option produces separate graphs for each shock variable, showing the IRFs with confidence intervals, as shown in Figure~\ref{fig.example5}. The SPJ results support MSV's main finding that household debt booms predict a boom--recession cycle in output.

\subsection{Example 6: Variance estimators}

The sixth example illustrates the \texttt{vce()} option. We return to the \citet{cerra2008growth} data of Example~4 and estimate the horizon-10 response under all supported variance estimators. The point estimates coincide across \texttt{vce()} choices, while only the standard errors differ. To save space, we display only the header and the coefficient on \texttt{CRISIS} for each command output.

\begin{stlog}
. use cs_f3data, clear
. xtset cnty obs
. xtlp cf10GRRT_WB CRISIS l1CRISIS l2CRISIS l3CRISIS l4CRISIS ///
>      l1GRRT_WB l2GRRT_WB l3GRRT_WB l4GRRT_WB ///
>          , fe m(spj) vce(unadjusted)

xtlp - SPJ - Individual Fixed Effects
(dropped 1 singleton observations)

Method: Split-panel jackknife                    Number of obs     =     2,806
Standard errors: Unadjusted
Small-sample adjustment: On

------------------------------------------------------------------------------
 cf10GRRT_WB | Coefficient  Std. err.      t    P>|t|     [95% conf. interval]
-------------+----------------------------------------------------------------
      CRISIS |     -7.893      0.969    -8.15   0.000       -9.792      -5.993
(output omitted)

. xtlp cf10GRRT_WB CRISIS l1CRISIS l2CRISIS l3CRISIS l4CRISIS ///
>      l1GRRT_WB l2GRRT_WB l3GRRT_WB l4GRRT_WB ///
>          , fe m(fe) vce(unadjusted)
(output omitted)
. xtreg cf10GRRT_WB CRISIS l1CRISIS l2CRISIS l3CRISIS l4CRISIS ///
>      l1GRRT_WB l2GRRT_WB l3GRRT_WB l4GRRT_WB ///
>      , fe
(output omitted)
. reghdfe cf10GRRT_WB CRISIS l1CRISIS l2CRISIS l3CRISIS l4CRISIS ///
>      l1GRRT_WB l2GRRT_WB l3GRRT_WB l4GRRT_WB ///
>      , absorb(cnty)
(output omitted)
\end{stlog}

In all previous examples, we did not specify \texttt{vce()}, so the default is \texttt{vce(unadjusted)}, which reports conventional homoskedastic standard errors. In this CS data set, the dependent variable \(\Delta_{h}y_{i,t+h}\) has already been manually constructed and saved as \texttt{cf10GRRT\_WB}, so we can use it directly here. With individual fixed effects, \texttt{xtlp} with \texttt{m(fe)} produces the same point estimate and standard error as \texttt{xtreg} with \texttt{fe}, and as \texttt{reghdfe} with \texttt{absorb(cnty)}.

\begin{stlog}
. xtlp cf10GRRT_WB CRISIS l1CRISIS l2CRISIS l3CRISIS l4CRISIS ///
>      l1GRRT_WB l2GRRT_WB l3GRRT_WB l4GRRT_WB ///
>      , fe m(spj) vce(robust)

xtlp - SPJ - Individual Fixed Effects
(dropped 1 singleton observations)

Method: Split-panel jackknife                    Number of obs     =     2,806
Standard errors: Robust
Small-sample adjustment: On

------------------------------------------------------------------------------
             |               Robust
 cf10GRRT_WB | Coefficient  std. err.      t    P>|t|     [95% conf. interval]
-------------+----------------------------------------------------------------
      CRISIS |     -7.893      1.008    -7.83   0.000       -9.870      -5.915
(output omitted)

. xtlp cf10GRRT_WB CRISIS l1CRISIS l2CRISIS l3CRISIS l4CRISIS ///
>      l1GRRT_WB l2GRRT_WB l3GRRT_WB l4GRRT_WB ///
>      , fe m(fe) vce(robust)
(output omitted)
. reghdfe cf10GRRT_WB CRISIS l1CRISIS l2CRISIS l3CRISIS l4CRISIS ///
>      l1GRRT_WB l2GRRT_WB l3GRRT_WB l4GRRT_WB ///
>      , absorb(cnty) vce(robust)
(output omitted)
\end{stlog}

With \texttt{vce(robust)}, the command reports heteroskedasticity-robust standard errors. Accordingly, the output labels the standard-error column as \texttt{Robust~std.~err.}. The point estimate and robust standard error for the FE method coincide with those obtained from \texttt{reghdfe} using \texttt{absorb(cnty)} and \texttt{vce(robust)}.

\begin{stlog}
. xtlp cf10GRRT_WB CRISIS l1CRISIS l2CRISIS l3CRISIS l4CRISIS ///
>      l1GRRT_WB l2GRRT_WB l3GRRT_WB l4GRRT_WB ///
>      , fe m(spj) vce(cluster cnty)

xtlp - SPJ - Individual Fixed Effects
(dropped 1 singleton observations)

Method: Split-panel jackknife                    Number of obs     =     2,806
Standard errors: One-way clustered
Small-sample adjustment: On

                                 (Std. err. adjusted for 137 clusters in cnty)
------------------------------------------------------------------------------
             |               Cluster
 cf10GRRT_WB | Coefficient  std. err.      t    P>|t|     [95% conf. interval]
-------------+----------------------------------------------------------------
      CRISIS |     -7.893      1.514    -5.21   0.000      -10.887      -4.898
(output omitted)

. xtlp cf10GRRT_WB CRISIS l1CRISIS l2CRISIS l3CRISIS l4CRISIS ///
>      l1GRRT_WB l2GRRT_WB l3GRRT_WB l4GRRT_WB ///
>      , fe m(fe) vce(cluster cnty)
(output omitted)
. xtreg cf10GRRT_WB CRISIS l1CRISIS l2CRISIS l3CRISIS l4CRISIS ///
>      l1GRRT_WB l2GRRT_WB l3GRRT_WB l4GRRT_WB ///
>      , fe cluster(cnty)
(output omitted)
. reghdfe cf10GRRT_WB CRISIS l1CRISIS l2CRISIS l3CRISIS l4CRISIS ///
>      l1GRRT_WB l2GRRT_WB l3GRRT_WB l4GRRT_WB ///
>      , absorb(cnty) cluster(cnty)
(output omitted)
\end{stlog}

With \texttt{vce(cluster cnty)}, the command reports standard errors clustered at the country level. The output therefore labels the standard-error column as \texttt{Cluster~std.~err.} and adds the note that the standard errors are adjusted for 137 clusters in \texttt{cnty}. The point estimate and clustered standard error for the FE method coincide with those obtained from \texttt{xtreg} using \texttt{fe} and \texttt{cluster(cnty)}, as well as from \texttt{reghdfe} using \texttt{absorb(cnty)} and \texttt{cluster(cnty)}.

\begin{stlog}
. xtlp cf10GRRT_WB CRISIS l1CRISIS l2CRISIS l3CRISIS l4CRISIS ///
>      l1GRRT_WB l2GRRT_WB l3GRRT_WB l4GRRT_WB ///
>      , fe m(spj) vce(cluster cnty obs)

xtlp - SPJ - Individual Fixed Effects
(dropped 1 singleton observations)
Warning: VCV matrix was non-positive semi-definite; adjustment from Cameron, Gelbac
> h & Miller applied. (h=0)

Method: Split-panel jackknife                    Number of obs     =     2,806
Standard errors: Two-way clustered
Small-sample adjustment: On

                              (Std. err. adjusted for 27 clusters in cnty obs)
------------------------------------------------------------------------------
             |               Cluster
 cf10GRRT_WB | Coefficient  std. err.      t    P>|t|     [95% conf. interval]
-------------+----------------------------------------------------------------
      CRISIS |     -7.893      1.698    -4.65   0.000      -11.382      -4.403
(output omitted)

. xtlp cf10GRRT_WB CRISIS l1CRISIS l2CRISIS l3CRISIS l4CRISIS ///
>      l1GRRT_WB l2GRRT_WB l3GRRT_WB l4GRRT_WB ///
>      , fe m(fe) vce(cluster cnty obs)
(output omitted)
. reghdfe cf10GRRT_WB CRISIS l1CRISIS l2CRISIS l3CRISIS l4CRISIS ///
>      l1GRRT_WB l2GRRT_WB l3GRRT_WB l4GRRT_WB ///
>      , absorb(cnty) cluster(cnty obs)
(output omitted)
\end{stlog}

With \texttt{vce(cluster cnty obs)}, the command reports two-way clustered standard errors, allowing arbitrary correlation within countries and within years. The output adds the note that the standard errors are adjusted for 27 clusters in \texttt{cnty} and \texttt{obs}. In this example, the estimated variance--covariance matrix is not positive semi-definite, so \texttt{xtlp} applies the \citet{cameron2011robust} adjustment, as does \texttt{reghdfe}. The point estimate and two-way clustered standard error for the FE method coincide with those obtained from \texttt{reghdfe} using \texttt{absorb(cnty)} and \texttt{cluster(cnty obs)}.

\begin{stlog}
. xtlp cf10GRRT_WB CRISIS l1CRISIS l2CRISIS l3CRISIS l4CRISIS ///
>      l1GRRT_WB l2GRRT_WB l3GRRT_WB l4GRRT_WB ///
>      , fe m(spj) vce(dkraay)

xtlp - SPJ - Individual Fixed Effects
(dropped 1 singleton observations)

Method: Split-panel jackknife                    Number of obs     =     2,806
Standard errors: Driscoll-Kraay
Maximum lag: 2
Small-sample adjustment: On

                                  (Std. err. adjusted for 27 clusters in obs)
------------------------------------------------------------------------------
             |             Drisc/Kraay
 cf10GRRT_WB | Coefficient  std. err.      t    P>|t|     [95% conf. interval]
-------------+----------------------------------------------------------------
      CRISIS |     -7.893      1.730    -4.56   0.000      -11.448      -4.338
(output omitted)

. xtlp cf10GRRT_WB CRISIS l1CRISIS l2CRISIS l3CRISIS l4CRISIS ///
>      l1GRRT_WB l2GRRT_WB l3GRRT_WB l4GRRT_WB ///
>      , fe m(fe) vce(dkraay)
(output omitted)
. reghdfe cf10GRRT_WB CRISIS l1CRISIS l2CRISIS l3CRISIS l4CRISIS ///
>      l1GRRT_WB l2GRRT_WB l3GRRT_WB l4GRRT_WB ///
>      , absorb(cnty) vce(dkraay)
(output omitted)
. xtlp cf10GRRT_WB CRISIS l1CRISIS l2CRISIS l3CRISIS l4CRISIS ///
>      l1GRRT_WB l2GRRT_WB l3GRRT_WB l4GRRT_WB ///
>      , fe m(fe) vce(dkraay lag(2))
(output omitted)
. reghdfe cf10GRRT_WB CRISIS l1CRISIS l2CRISIS l3CRISIS l4CRISIS ///
>      l1GRRT_WB l2GRRT_WB l3GRRT_WB l4GRRT_WB ///
>      , absorb(cnty) vce(dkraay 3)
(output omitted)

. xtlp cf10GRRT_WB CRISIS l1CRISIS l2CRISIS l3CRISIS l4CRISIS ///
>      l1GRRT_WB l2GRRT_WB l3GRRT_WB l4GRRT_WB ///
>      , fe m(spj) vce(dkraay, nodfadj)

xtlp - SPJ - Individual Fixed Effects
(dropped 1 singleton observations)

Method: Split-panel jackknife                    Number of obs     =     2,806
Standard errors: Driscoll-Kraay (nodfadj)
Maximum lag: 2
Small-sample adjustment: On

                                  (Std. err. adjusted for 27 clusters in obs)
------------------------------------------------------------------------------
             |             Drisc/Kraay
 cf10GRRT_WB | Coefficient  std. err.      t    P>|t|     [95% conf. interval]
-------------+----------------------------------------------------------------
      CRISIS |     -7.893      1.687    -4.68   0.000      -11.360      -4.425
(output omitted)

. xtlp cf10GRRT_WB CRISIS l1CRISIS l2CRISIS l3CRISIS l4CRISIS ///
>      l1GRRT_WB l2GRRT_WB l3GRRT_WB l4GRRT_WB ///
>      , fe m(fe) vce(dkraay, nodfadj)
(output omitted)
. xtscc cf10GRRT_WB CRISIS l1CRISIS l2CRISIS l3CRISIS l4CRISIS ///
>      l1GRRT_WB l2GRRT_WB l3GRRT_WB l4GRRT_WB ///
>      , fe
(output omitted)
. xtlp cf10GRRT_WB CRISIS l1CRISIS l2CRISIS l3CRISIS l4CRISIS ///
>      l1GRRT_WB l2GRRT_WB l3GRRT_WB l4GRRT_WB ///
>      , fe m(fe) vce(dkraay lag(2), nodfadj)
(output omitted)
. xtscc cf10GRRT_WB CRISIS l1CRISIS l2CRISIS l3CRISIS l4CRISIS ///
>      l1GRRT_WB l2GRRT_WB l3GRRT_WB l4GRRT_WB ///
>      , fe lag(2)
(output omitted)

. xtlp cf10GRRT_WB CRISIS l1CRISIS l2CRISIS l3CRISIS l4CRISIS ///
>      l1GRRT_WB l2GRRT_WB l3GRRT_WB l4GRRT_WB ///
>      , fe m(spj) vce(dkraay, ase)

xtlp - SPJ - Individual Fixed Effects
(dropped 1 singleton observations)

Method: Split-panel jackknife                    Number of obs     =     2,806
Standard errors: Driscoll-Kraay
Maximum lag: 2
Small-sample adjustment: Off

                                  (Std. err. adjusted for 27 clusters in obs)
------------------------------------------------------------------------------
             |             Drisc/Kraay
 cf10GRRT_WB | Coefficient  std. err.      t    P>|t|     [95% conf. interval]
-------------+----------------------------------------------------------------
      CRISIS |     -7.893      1.653    -4.78   0.000      -11.290      -4.495
(output omitted)

. xtlp cf10GRRT_WB CRISIS l1CRISIS l2CRISIS l3CRISIS l4CRISIS ///
>      l1GRRT_WB l2GRRT_WB l3GRRT_WB l4GRRT_WB ///
>      , fe m(fe) vce(dkraay, ase)
(output omitted)
. xtscc cf10GRRT_WB CRISIS l1CRISIS l2CRISIS l3CRISIS l4CRISIS ///
>      l1GRRT_WB l2GRRT_WB l3GRRT_WB l4GRRT_WB ///
>      , fe ase
(output omitted)
. xtlp cf10GRRT_WB CRISIS l1CRISIS l2CRISIS l3CRISIS l4CRISIS ///
>      l1GRRT_WB l2GRRT_WB l3GRRT_WB l4GRRT_WB ///
>      , fe m(fe) vce(dkraay lag(2), ase)
(output omitted)
. xtscc cf10GRRT_WB CRISIS l1CRISIS l2CRISIS l3CRISIS l4CRISIS ///
>      l1GRRT_WB l2GRRT_WB l3GRRT_WB l4GRRT_WB ///
>      , fe ase lag(2)
(output omitted)
\end{stlog}

With \texttt{vce(dkraay)}, the command reports Driscoll--Kraay standard errors. In this example, \texttt{xtlp} automatically selects a maximum lag of 2, with the small-sample adjustment turned on. The output labels the standard-error column as \texttt{Drisc/Kraay} and notes that the standard errors are adjusted for 27 clusters in \texttt{obs}. The resulting point estimate and Driscoll--Kraay standard error for the method FE coincide with those obtained from \texttt{reghdfe} using \texttt{absorb(cnty)} and \texttt{vce(dkraay)}. The syntax also allows users to specify the maximum lag explicitly through \texttt{vce(dkraay lag(2))}, which yields the same result as the default in this example. Note that \texttt{reghdfe} parameterizes the Driscoll--Kraay bandwidth as the maximum lag plus one; hence \texttt{vce(dkraay lag(2))} in \texttt{xtlp} corresponds to \texttt{vce(dkraay 3)} in \texttt{reghdfe}. 

By contrast, \texttt{vce(dkraay, nodfadj)} turns off the degrees-of-freedom adjustment inside the Driscoll--Kraay variance estimator and matches \texttt{xtscc} with \texttt{fe}. Finally, \texttt{vce(dkraay, ase)} turns off the small-sample adjustment for the reported standard errors; the header therefore reports \texttt{Small-sample adjustment:~Off}. This option is designed to align the output with \texttt{xtscc}, so \texttt{vce(dkraay, ase)} matches \texttt{xtscc} with \texttt{fe ase}. More generally, however, \texttt{ase} is available for all variance estimators in \texttt{xtlp}: it can be added after the comma inside \texttt{vce()} to turn off the small-sample adjustment. The same equivalences hold when the lag length is specified explicitly with \texttt{lag(2)}.

\subsection{Example 7: Singleton handling}

By default, \texttt{xtlp} removes singleton groups before estimation, which exhibit no within-group variation after absorbing the fixed effects and therefore do not contribute to identifying the within estimator. Keeping such observations can nevertheless affect the estimated variance--covariance matrix and may lead to downward-biased standard errors, as discussed by \citet{correia2015singletons}. For this reason, \texttt{xtlp} drops singleton groups by default. Users who want to retain them can disable this feature by specifying the \texttt{keepsingletons} option; in that case, \texttt{xtlp} displays a warning in red. This behavior is consistent with \texttt{reghdfe} when using \texttt{keepsingletons}.

\begin{stlog}
. use cs_f3data, clear
. xtset cnty obs

. xtlp cf3GRRT_WB CRISIS l1CRISIS l2CRISIS l3CRISIS l4CRISIS ///
>      l1GRRT_WB l2GRRT_WB l3GRRT_WB l4GRRT_WB ///
>      , fe m(spj) vce(robust) keepsin
WARNING: Singleton observations not dropped; statistical significance is biased

xtlp - SPJ - Individual Fixed Effects

Method: Split-panel jackknife                    Number of obs     =     3,808
Standard errors: Robust
Small-sample adjustment: On

------------------------------------------------------------------------------
             |               Robust
  cf3GRRT_WB | Coefficient  std. err.      t    P>|t|     [95% conf. interval]
-------------+----------------------------------------------------------------
      CRISIS |     -4.788      0.613    -7.80   0.000       -5.990      -3.585
(output omitted)

. xtlp cf3GRRT_WB CRISIS l1CRISIS l2CRISIS l3CRISIS l4CRISIS ///
>      l1GRRT_WB l2GRRT_WB l3GRRT_WB l4GRRT_WB ///
>      , fe m(fe) vce(robust) keepsin
(output omitted)
. reghdfe cf3GRRT_WB CRISIS l1CRISIS l2CRISIS l3CRISIS l4CRISIS ///
>      l1GRRT_WB l2GRRT_WB l3GRRT_WB l4GRRT_WB ///
>      , absorb(cnty) vce(robust) keepsingletons
(output omitted)
\end{stlog}

\section{Conclusion}

This paper introduces the new Stata command \texttt{xtlp} for estimating IRFs with panel LP. The command implements bias-corrected panel LP estimation using the SPJ and supports single- and multi-horizon estimation, flexible model specifications with alternative transformations of the dependent variable, automatic graphing, and a range of variance estimators. The SPJ implemented in \texttt{xtlp} removes the first-order bias, thereby restoring the validity of standard asymptotic inference. 
The command \texttt{xtlp} makes the panel LP method accessible to the broader Stata user community and supports credible empirical work using dynamic panel LP.

\bibliographystyle{sj-xtlp}
\bibliography{xtlp}

@article{jorda2005estimation,
  title={Estimation and inference of impulse responses by local projections},
  author={Jord{\`a}, {\`O}scar},
  journal={American Economic Review},
  volume={95},
  pages={161--182},
  year={2005}
}

@article{nickell1981biases,
 author = {Stephen Nickell},
 journal = {Econometrica},
 pages = {1417--1426},
 publisher = {Wiley, Econometric Society},
 title = {Biases in dynamic models with fixed effects},
 volume = {49},
 year = {1981}
}

@Article{Dhaene2015,
  author    = {Dhaene, Geert and Jochmans, Koen},
  journal   = {Review of Economic Studies},
  title     = {Split-panel jackknife estimation of fixed-effect models},
  year      = {2015},
  pages     = {991--1030},
  volume    = {82},
  publisher = {Oxford University Press},
}

@article{mei2026nickell,
  title={{Nickell bias in panel local projection: Financial crises are worse than you think}},
  author={Mei, Ziwei and Sheng, Liugang and Shi, Zhentao},
  journal={Journal of International Economics},
  volume={160},
  pages={104210},
  year={2026},
  publisher={Elsevier}
}

@article{romer2017new,
  title={New evidence on the aftermath of financial crises in advanced countries},
  author={Romer, Christina D and Romer, David H},
  journal={American Economic Review},
  volume={107},
  pages={3072--3118},
  year={2017},
  publisher={American Economic Association 2014 Broadway, Suite 305, Nashville, TN 37203}
}

@article{baron2021banking,
  title={Banking crises without panics},
  author={Baron, Matthew and Verner, Emil and Xiong, Wei},
  journal={Quarterly Journal of Economics},
  volume={136},
  pages={51--113},
  year={2021},
  publisher={Oxford University Press}
}

@article{mian2017household,
  title={Household debt and business cycles worldwide},
  author={Mian, Atif and Sufi, Amir and Verner, Emil},
  journal={Quarterly Journal of Economics},
  volume={132},
  pages={1755--1817},
  year={2017},
  publisher={Oxford University Press}
}

@article{cerra2008growth,
  title={Growth dynamics: The myth of economic recovery},
  author={Cerra, Valerie and Saxena, Sweta Chaman},
  journal={American Economic Review},
  volume={98},
  pages={439--457},
  year={2008},
  publisher={American Economic Association}
}

@article{sun2019xtspj,
  title={{\texttt{xtspj}}: A command for split-panel jackknife estimation},
  author={Sun, Yutao and Dhaene, Geert},
  journal={Stata Journal},
  volume={19},
  pages={335--374},
  year={2019},
  publisher={SAGE Publications Sage CA: Los Angeles, CA}
}

@misc{ugarte2023locproj,
  title={{\texttt{locproj}}: A new {Stata} command to estimate local projections},
  author={Ugarte-Ruiz, Alfonso},
  howpublished={Presentation at the 2023 Stata Conference, Stanford, CA, July 20--21},
  year={2023}
}

@article{acemoglu2019democracy,
  title={Democracy does cause growth},
  author={Acemoglu, Daron and Naidu, Suresh and Restrepo, Pascual and Robinson, James A},
  journal={Journal of Political Economy},
  volume={127},
  pages={47--100},
  year={2019},
  publisher={University of Chicago Press Chicago, IL}
}

@article{jorda2022longer,
  title={Longer-run economic consequences of pandemics},
  author={Jord{\`a}, {\`O}scar and Singh, Sanjay R and Taylor, Alan M},
  journal={Review of Economics and Statistics},
  volume={104},
  pages={166--175},
  year={2022},
  publisher={MIT Press One Rogers Street, Cambridge, MA 02142-1209, USA journals-info~…}
}

@article{gilchrist2022sovereign,
  title={Sovereign risk and financial risk},
  author={Gilchrist, Simon and Wei, Bin and Yue, Vivian Z and Zakraj{\v{s}}ek, Egon},
  journal={Journal of International Economics},
  volume={136},
  pages={103603},
  year={2022},
  publisher={Elsevier}
}

@article{hahn2002asymptotically,
  title={Asymptotically unbiased inference for a dynamic panel model with fixed effects when both $n$ and {$T$} are large},
  author={Hahn, Jinyong and Kuersteiner, Guido},
  journal={Econometrica},
  volume={70},
  pages={1639--1657},
  year={2002},
  publisher={JSTOR}
}

@article{driscoll1998consistent,
  title={Consistent covariance matrix estimation with spatially dependent panel data},
  author={Driscoll, John C and Kraay, Aart C},
  journal={Review of Economics and Statistics},
  volume={80},
  pages={549--560},
  year={1998},
  publisher={MIT Press 238 Main St., Suite 500, Cambridge, MA 02142-1046, USA journals~…}
}

@techreport{correia2015singletons,
  title={Singletons, cluster-robust standard errors and fixed effects: A bad mix},
  author={Correia, Sergio},
  type={Technical note},
  institution={Duke University},
  year={2015}
}

@article{cameron2011robust,
  title={Robust inference with multiway clustering},
  author={Cameron, A Colin and Gelbach, Jonah B and Miller, Douglas L},
  journal={Journal of Business \& Economic Statistics},
  volume={29},
  pages={238--249},
  year={2011},
  publisher={Taylor \& Francis}
}

@article{hoechle2007robust,
  title={Robust standard errors for panel regressions with cross-sectional dependence},
  author={Hoechle, Daniel},
  journal={Stata Journal},
  volume={7},
  pages={281--312},
  year={2007},
  publisher={SAGE Publications Sage CA: Los Angeles, CA}
}

@article{newey1994automatic,
  title={Automatic lag selection in covariance matrix estimation},
  author={Newey, Whitney K and West, Kenneth D},
  journal={Review of Economic Studies},
  volume={61},
  pages={631--653},
  year={1994},
  publisher={Wiley-Blackwell}
}

@misc{correia2017hdfe,
  title={Linear models with high-dimensional fixed effects: An efficient and feasible estimator},
  author={Correia, Sergio},
  howpublished={Working paper},
  year={2017}
}

@article{chudik2018half,
  title={Half-panel jackknife fixed-effects estimation of linear panels with weakly exogenous regressors},
  author={Chudik, Alexander and Pesaran, M Hashem and Yang, Jui-Chung},
  journal={Journal of Applied Econometrics},
  volume={33},
  pages={816--836},
  year={2018},
  publisher={Wiley Online Library}
}

@article{newey1987simple,
  title={A simple, positive semi-definite, heteroskedasticity and autocorrelation consistent covariance matrix},
  author={Newey, Whitney K and West, Kenneth D},
  journal={Econometrica},
  volume={55},
  pages={703--708},
  year={1987},
  publisher={JSTOR}
}

@Article{holtz1988estimating,
  author    = {Holtz-Eakin, Douglas and Newey, Whitney and Rosen, Harvey S},
  journal   = {Econometrica},
  title     = {Estimating vector autoregressions with panel data},
  year      = {1988},
  volume = {56},
  pages     = {1371--1395},
  publisher = {JSTOR},
}

@Book{Arellano2003,
  author    = {Arellano, Manuel},
  publisher = {Oxford University Press},
  title     = {Panel Data Econometrics},
  address   = {Oxford},
  year      = {2003},
}

@article{jorda2025local,
  title={Local projections},
  author={Jord{\`a}, {\`O}scar and Taylor, Alan M},
  journal={Journal of Economic Literature},
  volume={63},
  pages={59--110},
  year={2025}
}

\end{document}